\documentclass[journal]{IEEEtran}
\usepackage[latin9]{inputenc}
\usepackage{url}
\usepackage{amsmath}
\usepackage{amssymb}
\usepackage{graphicx}
\PassOptionsToPackage{normalem}{ulem}
\usepackage{ulem}

\makeatletter
\usepackage{amsfonts}
\usepackage{multirow}
\usepackage{url}
\providecommand{\U}[1]{\protect \rule{.1in}{.1in}}

\makeatother

\begin{document}
\title{Unified Simulation and Test Platform for Control Systems of Unmanned
Vehicles}
\author{Xunhua Dai, Chenxu Ke, Quan Quan and Kai-Yuan Cai \thanks{The authors are with School of Automation Science and Electrical Engineering,
Beihang University, Beijing 100191, China.}}
\maketitle
\begin{abstract}
Control systems on unmanned vehicles are safety-critical systems whose
requirements on reliability and safety are ever-increasing. Currently,
testing a complex autonomous control system is an expensive and time-consuming
process, which requires massive repeated experimental testing during
the whole development stage. This paper presents a unified simulation
and test platform for vehicle autonomous control systems aiming to
significantly improve the development speed and safety level of unmanned
vehicles. First, a unified modular modeling framework compatible with
different types of vehicles is proposed with methods to ensure modeling
credibility. Then, the simulation software system is developed by
the model-based design framework, whose modular programming methods
and automatic code generation functions ensure the efficiency, credibility,
and standardization of the system development process. Finally, an
FPGA-based real-time hardware-in-the-loop simulation platform is proposed
to ensure the comprehensiveness and credibility of the simulation
and test results. In the end, the proposed platform is applied to
a multicopter control system. By comparing with experimental results,
the accuracy and credibility of the simulation testing results are
verified by using the simulation credibility assessment method proposed
in our previous work. To verify the practicability of the proposed
platform, several successful applications are presented for the multicopter
rapid prototyping, estimation algorithm verification, autonomous flight
testing, and automatic safety testing with automatic fault injection
and result evaluation of unmanned vehicles.
\end{abstract}


\section{Introduction}

Unmanned vehicles (e.g., cars, boats, fixed-wing aircraft, multicopters,
robotics, and helicopters) are becoming increasingly popular in both
civil and military fields \cite{Nguyen2018}. For all unmanned vehicles,
safety is always the most basic requirement, and the concern over
potential safety issues remains the biggest challenge for the commercialization
of unmanned vehicles. For most commercial unmanned vehicles, there
is usually not enough space or payload to carry more hardware redundancy
(such as backup engines, actuators or motors) due to the limitation
of cost and performance, so the software redundancy is often applied
in control systems to ensure safety. As a result, among all components
on an unmanned vehicle, the control system is the most complex and
important component that undertakes responsibilities for both reliable
operation under nominal conditions and safety decision under failure
scenarios. Efficient simulation and test methods \cite{Goury2018}
for control systems of unmanned vehicles are urgently needed for the
ever-increasing system complexity and safety requirements.

\begin{figure}[tbh]
\centering \includegraphics[width=0.45\textwidth]{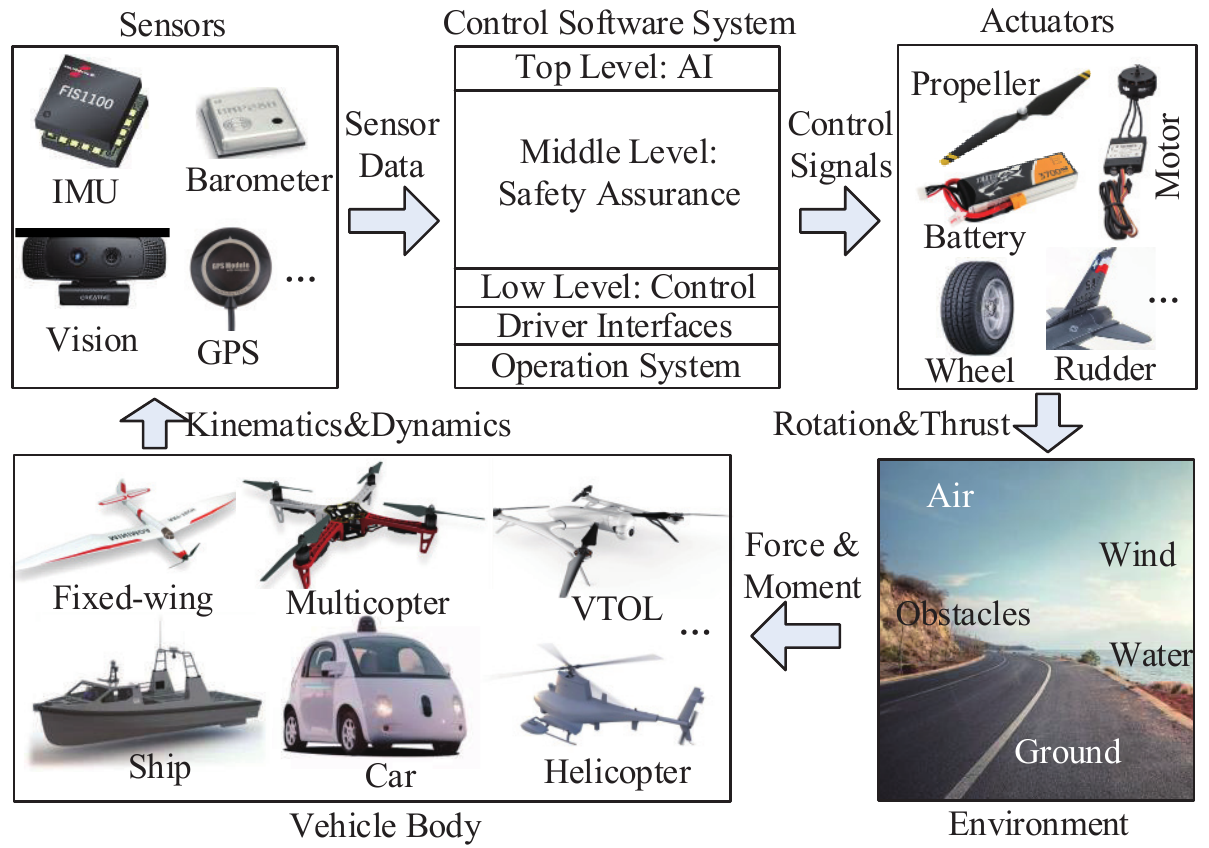}\caption{System structure of unmanned vehicles.}
\label{Fig01} 
\end{figure}

According to \cite{lipson2016driverless}, more than 80\% development
tasks of an autonomous control system are in the middle level (see
Fig.\,\ref{Fig01}) to guarantee safety under various possible faults.
However, most faults are rare to be encountered in practice, so massive
repeated experimental tests are essential to ensure that control systems
can correctly detect and handle unexpected faults. Although different
unmanned vehicles have different shapes, configurations, or running
environment, they have a similar system structure presented in Fig.\,\ref{Fig01}
and share many common model features and fault modes. The common faults
for unmanned vehicles include actuator faults (e.g., blocked, failed
or unhealthy), sensor and communication faults (e.g., loss of signal,
delays, GPS failed, and transmitting interference) \cite{Mebarki2015},
environment faults (e.g., obstacles, collisions, and wind disturbances)
\cite{Cole2018} and vehicle model faults (e.g., vibration and loss
of weight). Thus, a unified simulation and test platform compatible
with different types of vehicles will be beneficial to share fault
mode information and safety design experience to improve the safety
level of the whole unmanned vehicle field. Besides, it can help to
increase the exchange of safety design experience among different
companies, manufacturers and certification authorities, and decrease
the repetitive work during testing and assessment processes, which
is also beneficial to the rapid development requirements and better
response to the rules and regulations of governments.

Currently, experimental testing is widely adopted because it can reflect
real situations to the utmost extent. Besides, the safety problems
of control systems are usually highly coupled with the actual  situations.
Since there is still no widely recognized safety assessment standard
published for unmanned vehicle systems (both unmanned cars and aircraft),
many pre-researches are proposed to comprehensively test and assess
the safety of unmanned systems based on experimental methods. For
example, in \cite{Belcastro2017}, the experimental testing and assessing
method for the safety of control systems of Unmanned Aerial Vehicles
(UAVs) are studied inspired by the testing methods from the airworthiness
of manned aerial vehicles. However, for unmanned vehicles, the experimental
testing methods are usually high-cost, inefficient, dangerous, and
regulatory restricted. With the ever-increasing complexity of control
systems, the experimental testing methods become increasingly inefficient
in revealing potential safety issues and covering the testing cases.
Besides, in experimental testing, it is usually hard to obtain the
true states of unmanned vehicles (usually estimated by sensors or
human perceptions) and to ensure consistency of environment variables
(e.g., wind, temperature, and road condition). For example, the sensor
failures may cause the vehicle state estimation inaccuracy and unreliable
to reflect the true states of vehicles, so the test results in these
cases are not suitable for quantitative assessment of vehicle safety.
As a result, the experimental testing results usually need to be analyzed
by experienced engineers to evaluate the safety level of a control
system, which is not efficient, precise, and automatic enough for
modern complex control systems. Due to the above disadvantages of
experimental testing methods, new simulation and test methods (e.g.,
the real-time simulation methods \cite{Noureen2018}, high-precision
modeling and system identification methods \cite{tischler2018system},
model-based safety assessment methods \cite{Lisagor2011}) are becoming
the trend for both manned and unmanned vehicles. Although experimental
testing cannot be completely abandoned, simulation testing techniques
are taking on more and more safety testing and assessment tasks \cite{Jung2007}.

\begin{figure}[tbh]
\centering \includegraphics[width=0.45\textwidth]{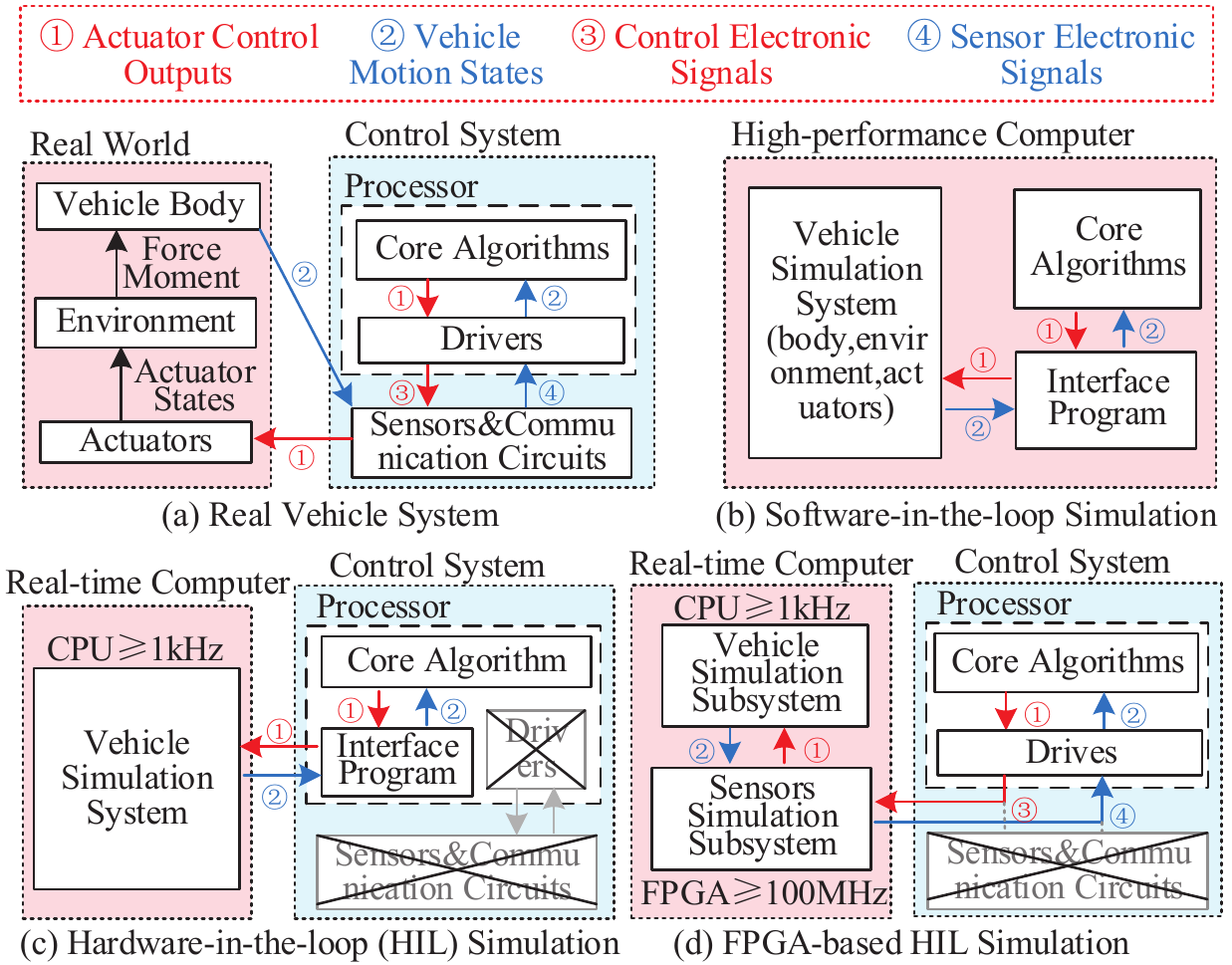}\caption{Comparisons between common simulation methods.}
\label{Fig02} 
\end{figure}

Simulation methods for vehicle control systems can be divided into
Software-In-the-Loop (SIL) simulation and Hardware-In-the-Loop (HIL)
simulation. As shown in Fig.\,\ref{Fig02}(b), by running the control
algorithms in the same computer with the vehicle simulation model,
SIL simulation can quickly test the control algorithms with the simulation
speed much faster than the real world. However, the software and hardware
operating environment of the SIL simulation is different from the
real vehicle system, whose control algorithms are running on specific
hardware (e.g., embedded systems, and industrial computers). In summary,
SIL simulations can accelerate the test speed, but it is based on
the expense of losing simulation credibility. To improve the simulation
credibility, as shown in Fig.\,\ref{Fig02}(c), HIL simulation is
proposed to reflect the real operating environment of the control
algorithms by using real control systems and real-time simulation
computers. The HIL simulation requires the simulation model runs in
real-time with the control systems, which makes it convenient to communicate
with other external hardware. Currently, there are many popular simulation
software supporting HIL simulations for accelerating the development
efficiency, such as \textit{Airsim}, \textit{SwarmSim}, and other
systems for UAVs \cite{Shah2018,worth2017swarmsim,Krichen2018}, and
\textit{Carsim}, \textit{Apollo}, and other systems for autonomous
cars \cite{sarhadi2015state,sylnice2018dynamic,chen2018autonomous}.
Besides, HIL simulation systems are also widely used in robotics studies
to verify their algorithms \cite{Franchi2012,Liu2018} or train the
autonomous vehicle control systems \cite{Lieaaw0863}. These HIL simulation
systems have been proven to be convenient and efficient in accelerating
the development speed of control systems of unmanned vehicles, but
there are still some problems. First, these HIL systems focus more
on providing testing environments for the upper-level algorithms such
as control algorithms, Computer Vision (CV) algorithms and Artificial
Intelligence (AI) algorithms of control systems; they usually cannot
simulate the low-level software (e.g., operating system, and drivers)
and hardware (e.g., sensor chips and high-speed analog circuits) due
to computing capability limitations. Secondly, they require to modify
the code of control systems to disable the original hardware drivers
and add interface programs to exchanging data with the simulation
computer, which may affect the operating environment and performance
of the original system (unstable and unreliable). Thirdly, the lack
of widely recognized assessment methods makes it hard for people (e.g.,
users, manufacturers, and certification authorities) to believe the
credibility of the simulation testing results for assessments and
certifications.

In the past, limited by the real-time response performance of simulation
computers, it is hard for Central Processing Units (CPUs) \cite{saad2015real}
on simulation computers to simulate sensors or other electronic chips
with high-speed interfaces or high-frequency analog circuits. For
example, a nanosecond-level real-time update frequency (100MHz) is
required to reliably simulate a sensor chip with the high-speed Serial
Peripheral Interface (SPI), which is a difficult task for traditional
CPU-based Commercial-Off-The-Shelf (COTS) simulation computers whose
reliable update frequencies are usually within microsecond level.
In recent years, with the utilization of Field Programmable Gate Array
(FPGA) \cite{saad2015real}, COTS FPGA-based simulation computers
(e.g., OPAL-RT$^{\circledR}$/OP series and NI$^{\circledR}$/PXI
series) start to have the simulation performance of nanosecond-level
real-time update frequency \cite{Noureen2018,Mikkili2015}. This makes
it possible to simulate almost everything (including vehicle motion,
sensor chips, electric circuits, and high-speed interfaces) outside
the main processor of the control systems as presented in Fig.\,\ref{Fig02}(d).
Besides, the tested control systems can be treated as black boxes
in HIL simulation systems with no need for accessing the code or adding
interface programs. Therefore, by simply replacing the sensor models,
the HIL simulation system can also be applied to perform comprehensive
tests for different brands of control systems.

For all simulation methods, the primary challenge is to ensure simulation
credibility \cite{mehta2016simulation}, namely making people believe
the simulation results are as real as experiments in the real world.
The simulation credibility (compared with real systems) mainly determined
by two aspects: platform credibility and model credibility. The platform
credibility can be further divided into hardware credibility and software
credibility. As previously mentioned, the hardware credibility can
be guaranteed by the FPGA-based HIL simulation method, while the software
credibility and the model credibility are still challenges for simulation
methods. The Model-Based Design (MBD) \cite{MBD2018} method is an
effective means to solve the above credibility problems by using modular
visual programming technology and automatic code generation technology
to standardize the modeling, developing and testing procedures of
complex control systems. The MBD method can ensure the software credibility
by eliminating the disturbance factors such as manual programming
negligence and nonstandard development process. For example, the MathWorks$^{\circledR}$/MATLAB
(the most widely used MBD software) can ensure the generated code
meet the requirements of standards and guidelines such as DO-178C.
In MBD methods, the whole simulation systems can be divided into many
small subsystems (modules), such as kinematic modules, GPS modules,
ground modules, and propeller modules. Certification authorities can
verify and validate these modules to build a standard product model
database for companies to develop the vehicle prototype and the corresponding
vehicle simulation system. Then, the model credibility can be guaranteed
by using well-validated standard component models. What is more, with
MBD method, the same component model can be applied to different types
of vehicles and control systems, which may significantly improve the
design, verification, validation, and certification process of the
unmanned vehicle filed. For ensuring model credibility, in our previous
research \cite{Dai2019Simulation}, a credibility assessment method
is proposed based on experiments to assess the model credibility of
HIL simulation platforms from multiple aspects, such as key performance
indices, time-domain characteristics, and the frequency-domain characteristics.
By combining the above methods, the credibility of the simulation
platform can be guaranteed from the model, development process, and
platform hardware aspects.

In this paper, a unified simulation and test platform is proposed
for control systems of unmanned vehicles based on FPGA-based HIL simulation
and MBD methods, aiming to significantly improve the test efficiency
and safety level of control systems on unmanned vehicles. The main
research contents and the corresponding contributions are listed as
follows.

\textit{(i) Unified Modeling Method.} There are so many in common
among different types of unmanned vehicles. They should not be treated
separately as more and more composite vehicles (e.g., multicopter
+ fixed-wing and car + fixed-wing) emerged. Therefore, a unified modeling
method is proposed for all types of unmanned vehicles along with parameter
measurement and identification methods to validate the obtained model
and ensure simulation credibility. The corresponding content is presented
in \textit{Section \ref{sec:2}}.

\textit{(ii) Real-time HIL Test Platform with MBD.} A real-time HIL
test platform is built for the testing and assessment of control systems.
The platform is capable of simulating any real-world situation outside
the control software with advantages in obtaining the true states
and controlling the testing variables. The utilization of MBD methods
can ensure that the testing results are credible and standard-compliant.
Besides, the HIL simulation method and MBD method ensure the platform
can be easily applied to different brands of control systems, which
is convenient for both companies and manufacturers. The corresponding
content is presented in \textit{Section \ref{sec:3}}.

The significant advantages of the proposed unified HIL simulation
test platform are reflected in four aspects: extensibility, comprehensiveness,
verification, standardization.

\textbf{(i) Extensibility}. By changing the parameters (e.g., weight,
size, and aerodynamic coefficients) of specific subsystem modules,
it is easy to extend a simulation system to other vehicle systems
with similar structures. Moreover, by replacing a whole subsystem
module (e.g., a propeller module to a tire module), the vehicle simulation
system can be extended to other types of vehicles.

\textbf{(ii) Comprehensiveness}. The current simulation systems mainly
focus on functional testing, i.e., whether the vehicles can work properly
in normal situations. However, unmanned vehicles are safety-critical
systems, and most of the effects are focused on safety testing, i.e.,
whether the vehicles can work safely when accident and faults happen.
With the modular programming method, the fault modes, the aging process,
and the probabilistic reliability property can be modeled for each
subsystem module to improve the comprehensiveness of the simulation
platform. Mathematically, the fault injection simulations (or other
safety simulations) can be realized by online changing the module
parameter or functional expressions of a subsystem module while the
simulation program is running.

\textbf{(iii) Verification}. In practice, it is difficult to verify
and validate the simulation accuracy and credibility of a complex
simulation system. However, it is relatively simple to verify a small
subsystem. Therefore, the modular programming method can divide a
complex simulation system into many small subsystems, and verify it
from lower levels to the top level. More importantly, if all subsystems
used in a simulation system are well-verified modules from certification
authorities, the verification efficiency can be significantly improved.

\begin{figure}
\centering \includegraphics[width=0.45\textwidth]{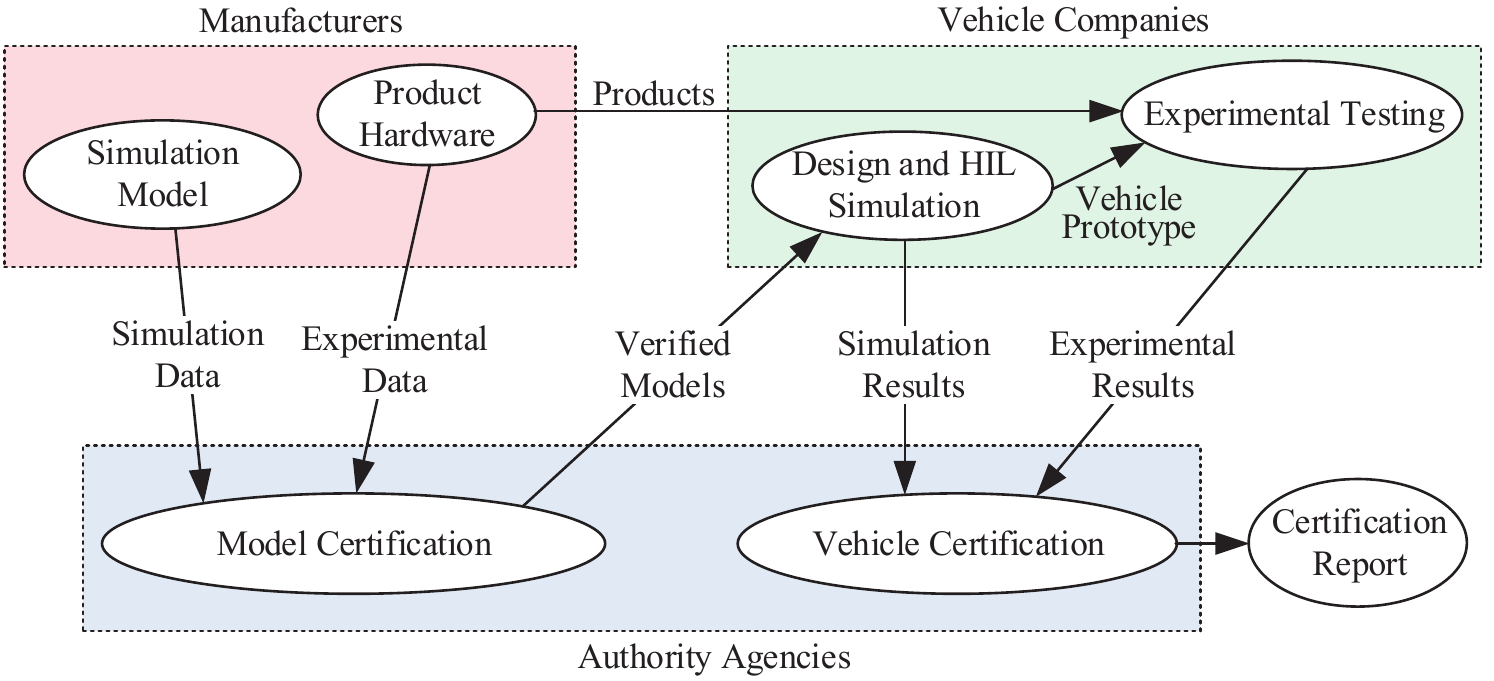}\caption{Certification framework for unmanned vehicles.}
\label{Fig04}
\end{figure}

\textbf{(iv) Standardization}. A standard certification framework
is urgently needed for unmanned vehicles to improve the testing and
certification efficiency. The modular programming method is a feasible
way to solve this problem with the certification framework presented
in Fig.\,\ref{Fig04}. In this framework, the manufactures should
provide the product hardware along with a simulation model which should
be fully verified and certificated by authority agencies based on
the simulation data and experimental data. That coincides with the
idea of Digital Twin \cite{Alexander2018Towards} for the efficient
design and testing of complex systems. Then, the vehicle companies
can use the certified models for simulation system development and
prototype design. Finally, the simulation results and experimental
results can be applied for the certification of the unmanned vehicle.

In \textit{Section \ref{sec:4}}, the successful application of the
testing platform on multicopter vehicles with the \textit{PX4} autopilot
system indicates the effectiveness of the proposed modeling and simulation
methods. To verify the practicability of the proposed platform, several
successful applications are presented for the multicopter rapid prototyping,
estimation algorithm verification, autonomous flight testing, and
automatic safety testing with automatic fault injection and result
evaluation of unmanned vehicles. In the end, \textit{Section \ref{sec:5}}
presents the conclusions.

\section{Unified Modeling Method}

\label{sec:2}
\begin{figure*}
\centering \includegraphics[width=0.7\textwidth]{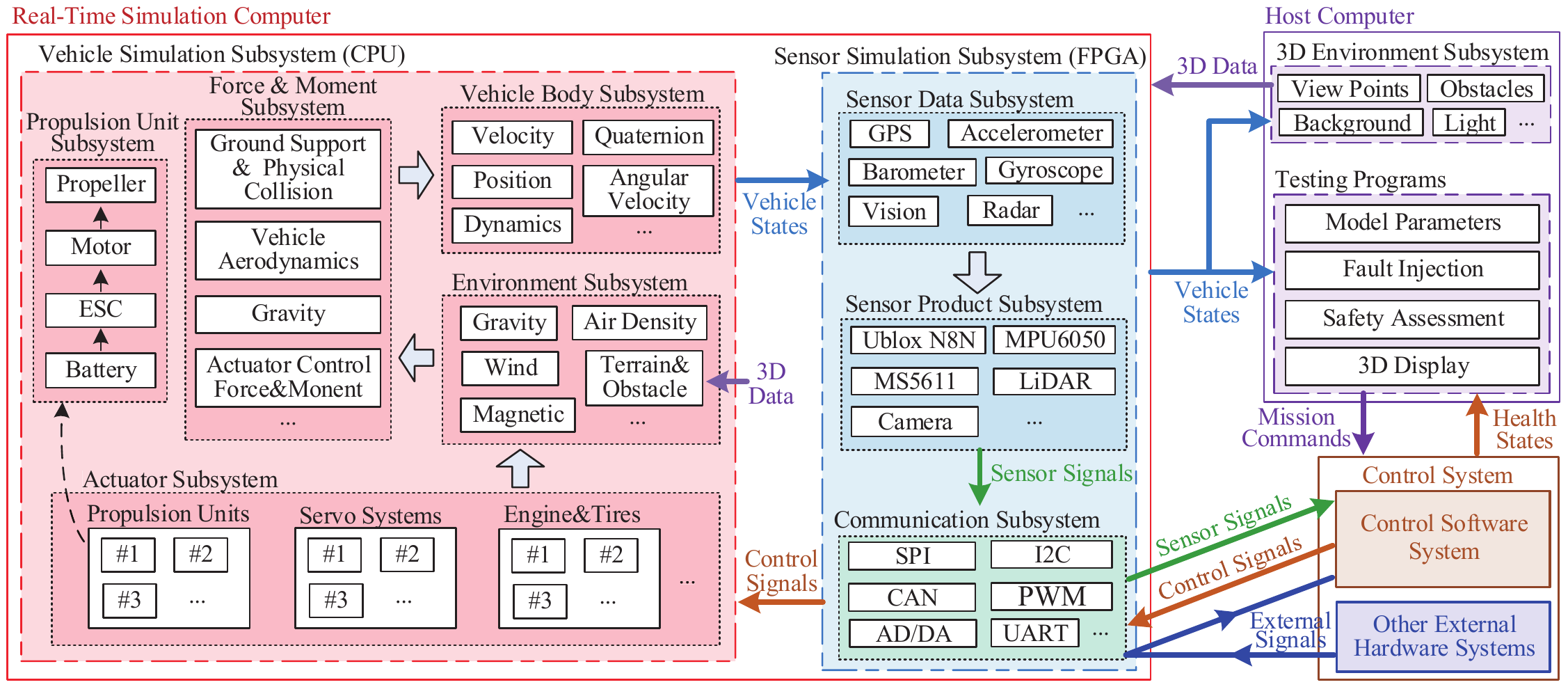}\caption{System structure of the simulation test platform for unmanned vehicles.}
\label{Fig03}
\end{figure*}

As shown in Fig.\,\ref{Fig01}, different types of
vehicles (e.g., cars, aircraft, and boats) have many common features
in modeling and simulation. To make the maximum utilization of these
common features, a unified simulation testing system is developed
with the system structure presented in Fig. \ref{Fig03} for all unmanned
vehicle systems. In this section, the unified modeling methods for
the simulation system in Fig. \ref{Fig03} will be introduced in detail.
The hardware structure and development process for the simulation
system in Fig. \ref{Fig03} will be introduced in Section \ref{sec:3}.

\subsection{Overall Vehicle Model}

\subsubsection{Model Abstraction}

In practice, a complex dynamic system usually consists of many small
subsystem systems (e.g., body system and propulsion systems). Every
subsystem includes input signals $\mathbf{u}_{*}$, output signals
$\mathbf{y}_{*}$, parameters $\mathbf{\Phi}_{*}$, dynamic states
$\mathbf{x}_{*}$, and and dynamic state and output functions $\mathbf{f}_{*}\left(\cdot\right)$
and $\mathbf{h_{*}}\left(\cdot\right)$, which can be mathematically
described by the following dynamic equations
\begin{equation}
\left\{ \begin{array}{c}
\mathbf{\dot{x}}_{*}=\mathbf{f}_{*}\left(\mathbf{x}_{*},\mathbf{\Phi}_{*},\mathbf{u}_{*}\right)\\
\mathbf{y}_{*}=\mathbf{h_{*}}\left(\mathbf{x}_{*},\mathbf{\Phi}_{*},\mathbf{u}_{*}\right)
\end{array}.\right.\label{eq:SysComp}
\end{equation}
For simplicity, a dynamic subsystem in Eq.\,(\ref{eq:SysComp}) is
further simplified to the following input/output form as
\begin{equation}
\mathbf{y}_{*}=\mathbf{S}_{*}\left(\mathbf{u}_{*}\right).\label{eq:SysSimp}
\end{equation}
The system description form in Eq.\,\ref{eq:SysSimp} will be applied
in the following content to describe the connection relationships
among different subsystems. The subscript symbol ``$*$'' in Eqs.\,(\ref{eq:SysComp})(\ref{eq:SysSimp})
can be replaced by different abbreviation words to represent different
dynamic systems, such as the control software system $\mathbf{S}_{\text{ctrl}}$
and the sensor simulation subsystem $\mathbf{S}_{\text{sens}}$. 

\subsubsection{Main Framework of Simulation System}

As presented in Fig. \ref{Fig03}, the whole simulation system can
be divided into three main subsystems: the vehicle simulation subsystem
$\mathbf{S}_{\text{vehi}}$ (generating vehicle states according to
the control signals), the 3D environment simulation subsystem $\mathbf{S}_{\text{3d}}$
(generating vision data according to the vehicle states), and the
sensor simulation subsystem $\mathbf{S}_{\text{sens}}$ (generating
sensor signals according to the vehicle states and vision data). Besides,
there is a control system $\mathbf{S}_{\text{ctrl}}$ (generating
control signals according to the sensor data) to be tested. The above
connection relationships among the above four systems are mathematically
described by
\begin{equation}
\begin{array}{l}
\mathbf{y}_{\text{ctrl}}=\mathbf{S}_{\text{ctrl}}\left(\mathbf{u}_{\text{ctrl}}\right),\,\mathbf{u}_{\text{ctrl}}=\mathbf{y}_{\text{sens}}\\
\mathbf{y}_{\text{vehi}}=\mathbf{S}_{\text{vehi}}\left(\mathbf{u}_{\text{vehi}}\right),\,\mathbf{u}_{\text{vehi}}=\left\{ \mathbf{y}_{\text{ctrl}},\mathbf{y}_{\text{3d}}\right\} \\
\mathbf{y}_{\text{3d}}=\mathbf{S}_{\text{3d}}\left(\mathbf{u}_{\text{3d}}\right),\,\mathbf{u}_{\text{3d}}=\mathbf{y}_{\text{vehi}}\\
\mathbf{y}_{\text{sens}}=\mathbf{S}_{\text{sens}}\left(\mathbf{u}_{\text{sens}}\right),\,\mathbf{u}_{\text{sens}}=\left\{ \mathbf{y}_{\text{vehi}},\mathbf{y}_{\text{3d}}\right\} 
\end{array}\label{eq:SysStru}
\end{equation}
which is consistent with the connection relationships in Fig. \ref{Fig03}.
In the following, the unified modeling methods for the above three
main subsystems will be introduced sequentially.

\subsection{Vehicle Simulation Subsystem}

\label{subsec:Vehicle-Simulation-Subsystem}

\begin{figure}
\centering \includegraphics[width=0.45\textwidth]{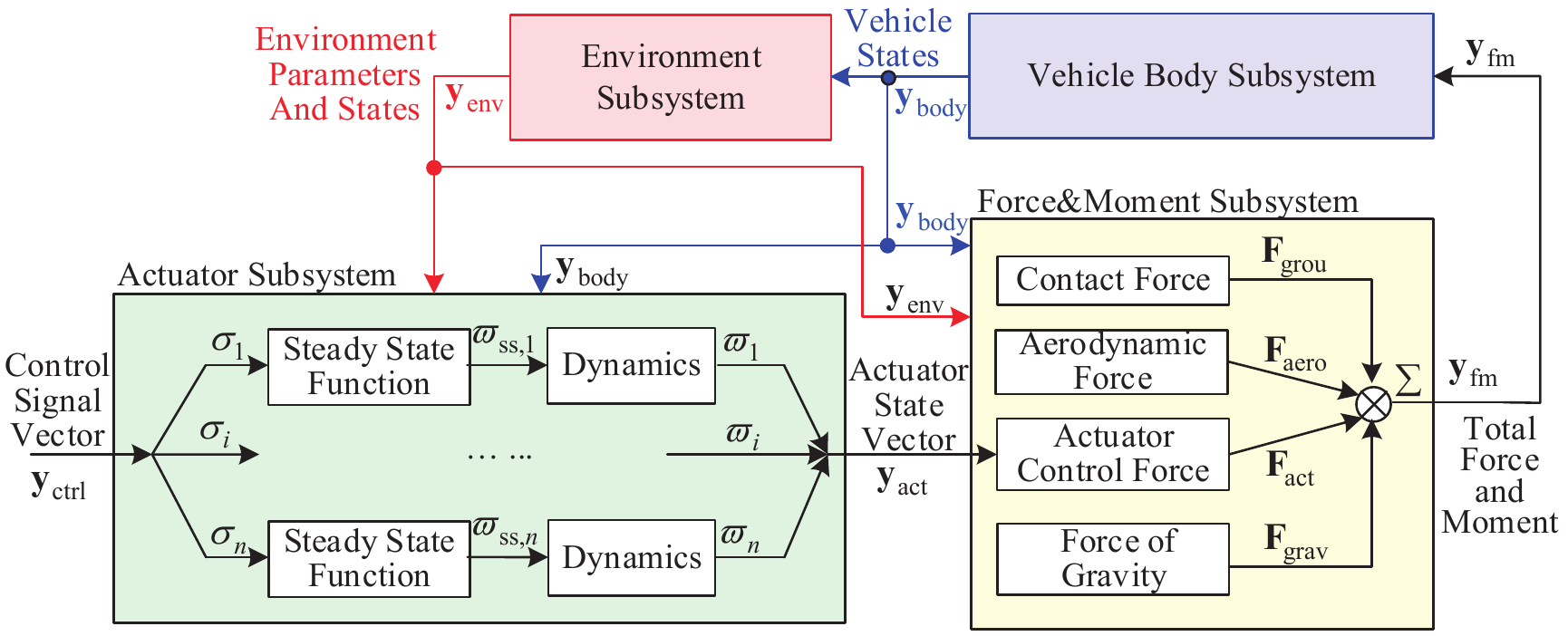}\caption{The structure of the vehicle simulation subsystem.}

\label{Fig05-1}
\end{figure}

The vehicle simulation subsystem $\mathbf{S}_{\text{vehi}}$ in Fig.
\ref{Fig03} can be further divided into four main subsystems: the
actuator subsystem $\mathbf{S}_{\text{act}}$, the environment subsystem
$\mathbf{S}_{\text{env}}$, the force\&moment subsystem $\mathbf{S}_{\text{fm}}$,
and the vehicle body subsystem $\mathbf{S}_{\text{body}}$. As shown
in Fig.\,\ref{Fig05-1}, the connection relationships of the four
subsystems are mathematically described as

\begin{equation}
\begin{array}{l}
\mathbf{y}_{\text{body}}=\mathbf{S}_{\text{body}}\left(\mathbf{u}_{\text{ctrl}}\right),\,\mathbf{u}_{\text{ctrl}}=\mathbf{y}_{\text{fm}}\\
\mathbf{y}_{\text{fm}}=\mathbf{S}_{\text{fm}}\left(\mathbf{u}_{\text{fm}}\right),\,\mathbf{u}_{\text{fm}}=\text{\ensuremath{\left\{  \mathbf{y}_{\text{body}},\mathbf{y}_{\text{act}},\mathbf{y}_{\text{env}}\right\} } }\\
\mathbf{y}_{\text{env}}=\mathbf{S}_{\text{env}}\left(\mathbf{u}_{\text{env}}\right),\,\mathbf{u}_{\text{env}}=\left\{ \mathbf{y}_{\text{body}},\mathbf{y}_{\text{3d}}\right\} \,\\
\mathbf{y}_{\text{act}}=\mathbf{S}_{\text{act}}\left(\mathbf{u}_{\text{act}}\right),\,\mathbf{u}_{\text{act}}=\left\{ \mathbf{y}_{\text{ctrl}},\mathbf{y}_{\text{body}},\mathbf{y}_{\text{env}}\right\} 
\end{array}\label{eq:VehicSys}
\end{equation}
where $\mathbf{y}_{\text{body}}$ contains vehicle motion states (e.g.,
position, velocity, and attitude), $\mathbf{y}_{\text{fm}}$ denotes
all the forces and moments acting on the vehicle, $\mathbf{y}_{\text{env}}$
includes environment parameters (e.g., gravity, air density, terrain,
and obstacle distribution) , $\mathbf{y}_{\text{act}}$ denotes actuator
states (e.g., rotating speed of rotors, and deflection angle of control
surface). By combining the output signals in Eq.\,(\ref{eq:VehicSys}),
the output set $\mathbf{y_{\text{vehi}}}$ for the vehicle simulation
subsystem $\mathbf{S}_{\text{vehi}}$ is given by
\[
\mathbf{y_{\text{vehi}}}\triangleq\left\{ \mathbf{y}_{\text{body}},\mathbf{y}_{\text{fm}},\mathbf{y}_{\text{env}},\mathbf{y}_{\text{act}}\right\} .
\]
The key modeling methods for the four subsystems in Eq. (\ref{eq:VehicSys})
will be introduced as follows.

\subsubsection{Vehicle body subsystem}

As shown in Fig.\,\ref{Fig05-1}, the vehicle body subsystem $\mathbf{S}_{\text{body}}$
computes the vehicle states $\mathbf{y}_{\text{body}}$ according
to the force and moment $\mathbf{y}_{\text{fm}}$ acting on the vehicle.
In practice, based on the flat-earth assumption (ignoring the curvature
of the earth in a small range) and the rigid-body assumption (the
body is rigid and not flexible), most vehicle body subsystem $\mathbf{S}_{\text{body}}$
can be described by Six-Degree-of-Freedom (6-DOF) equations \cite[pp. 25-54]{Stevens2004AirContrl},\cite[pp. 99-143]{quan2017introduction}
\begin{equation}
\left\{ \begin{array}{l}
^{\text{e}}\mathbf{\dot{p}}=\mathbf{R_{\text{b}}^{\text{e}}}\cdot\mathbf{^{\text{b}}v}\\
\mathbf{^{\text{b}}\dot{v}}=-\mathbf{^{\text{b}}\boldsymbol{\omega}}\times\mathbf{^{\text{b}}v}+\left(1/m\right)\cdot\mathbf{^{\text{b}}F}\\
\mathbf{\dot{R}_{\text{b}}^{\text{e}}}=\mathbf{R_{\text{b}}^{\text{e}}}\cdot\left[^{\text{b}}\boldsymbol{\omega}\right]_{\times}\\
\mathbf{J}\cdot{}^{\text{b}}\dot{\boldsymbol{\omega}}=-^{\text{b}}\boldsymbol{\omega}\times\left(\mathbf{J}\cdot{}^{\text{b}}\boldsymbol{\omega}\right)+\mathbf{^{\text{b}}M}
\end{array}\right.\label{eq:bodyEq}
\end{equation}
where the right-superscript symbols ``e'' and ``b'' denote the
North-East-Down (NED) earth frame and the Head-Right-Down body frame
\cite[pp. 25-54]{Stevens2004AirContrl} respectively; $^{\text{e}}\mathbf{p}\in\mathbb{R}^{3}$
is the position vector defined in the earth frame; $\mathbf{^{\text{b}}v}\in\mathbb{R}^{3}$
and $\mathbf{^{\text{b}}\boldsymbol{\omega}}\in\mathbb{R}^{3}$ are
the velocity vector and the angular velocity vector defined in the
body frame; $\mathbf{^{\text{b}}F\in\mathbb{R}^{\mathnormal{3}}}$
and $\mathbf{\mathbf{^{\text{b}}M}\in\mathbb{R}^{\mathnormal{3}}}$
are the force vector and the moment vector defined in the body frame;
$\mathbf{R_{\text{b}}^{\text{e}}}\in\mathbb{R}^{3\times3}$ is the
rotation matrix to transform a vector from body frame to earth frame;
$\mathbf{J\in\mathbb{R}^{\text{\ensuremath{3\times3}}}}$ and $m$
are the moment of inertia matrix and the mass of the vehicle. The
6-DOF dynamic equations in Eq.\,(\ref{eq:bodyEq}) can be applied
to describe the vehicle body module $\mathbf{y}_{\text{body}}=\mathbf{S}_{\text{body}}\left(\mathbf{y}_{\text{fm}}\right)$
in Eq.\,(\ref{eq:SysComp}), where the state set is $\mathbf{x}_{\text{body}}\triangleq\left\{ ^{\text{e}}\mathbf{p},\mathbf{^{\text{b}}v},\mathbf{R_{\text{b}}^{\text{e}}},{}^{\text{b}}\mathbf{w}\right\} $,
the input set is $\mathbf{u_{\text{body}}}=\mathbf{y}_{\text{fm}}\triangleq\left\{ \mathbf{^{\text{b}}F},\mathbf{^{\text{b}}M}\right\} $,
the parameter set is $\mathbf{\Phi}_{\text{body}}\triangleq\left\{ \mathbf{J},m\right\} $,
and the output set is $\mathbf{y}_{\text{body}}\triangleq\left\{ \mathbf{x}_{\text{body}},\mathbf{^{\text{e}}v},{}^{\text{b}}\mathbf{\dot{v}},{}^{\text{b}}\boldsymbol{\omega},\mathbf{R_{\text{b}}^{\text{e}}},\cdots\right\} $.

\subsubsection{Environment subsystem}

As shown in Fig.\,\ref{Fig05-1}, the environment subsystem $\mathbf{S}_{\text{env}}$
generates environment parameters $\mathbf{y}_{\text{env}}$ (e.g.,
air density, temperature, terrain, wind, and magnetic field) based
on the position of the vehicle $^{\text{e}}\mathbf{p}\in\mathbf{y}_{\text{body}}$.
In practice, the World Geodetic System (WGS84) model \cite{smith1987department}
is widely used to describe the shape of the earth, which can convert
the position vector $^{\text{e}}\mathbf{p}$ to the earth Latitude-Longitude-Altitude
(LLA) global position $^{\text{e}}\mathbf{p}_{\text{g}}\triangleq\left[\mu\,\iota\,h\right]^{\text{T}}$,
where $\mu,\iota$ (unit: degree) are the latitude and longitude,
and $h$ (unit: m) is the altitude. Then, the acceleration of gravity
$g$ can be estimated by the WGS model \cite{smith1987department}
based on the vehicle global position $^{\text{e}}\mathbf{p}_{\text{g}}$.
Similarly, the air density and temperature are estimated by the International
Standard Atmosphere (ISA) model \cite{cavcar2000international}, and
the magnetic field vector is estimated by the World Magnetic Model
(WMM) \cite{chulliat2015us}. Besides, according to the Military Specification
MIL-F-8785C \cite{moorhouse1982background}, the wind velocity disturbance
vector $^{\text{e}}\mathbf{v}_{\text{wind}}\in\mathbb{R}^{3}$ (defined
in the earth frame) can be described by the following superposition
form
\begin{equation}
^{\text{e}}\mathbf{v}_{\text{wind}}={}^{\text{e}}\mathbf{v}_{\text{turb}}+{}^{\text{e}}\mathbf{v}_{\text{cons}}+{}^{\text{e}}\mathbf{v}_{\text{sheer}}+{}^{\text{e}}\mathbf{v}_{\text{gust}}\label{eq:Wind}
\end{equation}
where $^{\text{e}}\mathbf{v}_{\text{turb}}$ denotes the atmospheric
turbulence field, $^{\text{e}}\mathbf{v}_{\text{\text{cons}}}$ denotes
the prevailing wind field, $^{\text{e}}\mathbf{v}_{\text{sheer}}$
denotes the wind shear field, and $^{\text{e}}\mathbf{v}_{\text{gust}}$
denotes the wind gust field. There many widely used mathematical models
for the wind components in Eq.\,(\ref{eq:Wind}). For example, the
wind turbulence $^{\text{e}}\mathbf{v}_{\text{turb}}$ can be described
by the Dryden Wind Turbulence Model \cite{moorhouse1982background}.

\subsubsection{Actuator subsystem}

As shown in Fig.\,\ref{Fig05-1}, the actuator subsystem $\mathbf{S}_{\text{act}}$
outputs actuator state $\mathbf{y}_{\text{act}}$ according to the
control input $\mathbf{y}_{\text{ctrl}}$ from the control system.
In practice, it is difficult to obtain the mathematical model of an
actuator system because it is usually composed by complex mechanical
components along with programmable control units, such as the Electronic
Speed Controller (ESC) for UAV brushless motors, and the Electronic
Control Unit (ECU) for car engines and steering systems. These control
units have feedback control to ensure that the actuator steady output
$\delta_{\text{ss},i}$ satisfy the preprogrammed function of the
input control signal $\sigma_{i}\in\mathbf{y}_{\text{ctrl}}$ under
different operating environment. According to \cite{quan2017introduction},
a complex actuator system can be linearized to a steady-state process
$f_{\text{ss},i}\left(\cdot\right)$ and a dynamic response process
$G_{\text{ss},i}\left(s\right)$ around the rated operation condition.
For example, a motor-propeller system with an ESC can be simplified
as a first-order or second-order inertial process $G_{\text{ss},i}\left(s\right)$
and a steady-state function $f_{\text{ss},i}\left(\sigma_{i}\right)$
as
\begin{equation}
\delta_{i}=G_{\text{ss},i}\left(s\right)\cdot f_{\text{ss},i}\left(\sigma_{i}\right).\label{eq:GS-1}
\end{equation}
Noteworthy, $f_{\text{ss},i}\left(\cdot\right)$ can be measured by
static testing, and $G_{\text{ss},i}\left(s\right)$ can be measured
by system identification methods through frequency-response testing
\cite{remple2006aircraft}. By using Eq.\,(\ref{eq:GS-1}), it is
easy to obtain the actuator output signal $\delta_{i}\left(t\right)$
(propeller rotating speed) under the given control signal $\sigma_{i}\left(t\right)$
(throttle control signal). Then, the control force and torque generated
by the state of an actuator $\delta_{i}$ can be obtained by the ground
friction model, aerodynamic model, or other mechanical models \cite{quan2017introduction,cai2011unmanned,rajamani2011vehicle}.

\subsubsection{Force \& moment subsystem}

As shown in Fig.\,\ref{Fig05-1}, the force\&moment subsystem $\mathbf{S}_{\text{fm}}$
outputs the force and moment $\mathbf{y}_{\text{fm}}\triangleq\left\{ \mathbf{^{\text{b}}F},\mathbf{^{\text{b}}M}\right\} $
to the vehicle body subsystem $\mathbf{S}_{\text{body}}$. The total
force $\mathbf{^{\text{b}}F}$ and moment $\mathbf{^{\text{b}}M}$
acting on a vehicle can be divided into many components from different
sources. Taking the force vector $\mathbf{^{\text{b}}F}\in\mathbf{y}_{\text{fm}}$
(defined in the body frame) as an example, it can be described by
the following superposition form

\begin{equation}
\mathbf{^{\text{b}}F}=\mathbf{^{\text{b}}F}_{\text{aero}}+\mathbf{^{\text{b}}F}_{\text{grav}}+\mathbf{^{\text{b}}F}_{\text{cont}}+\sum\mathbf{^{\text{b}}F}_{\text{act},i}\label{eq:totF}
\end{equation}
where $\mathbf{^{\text{b}}F}_{\text{aero}}\in\mathbb{R}^{\mathnormal{3}}$
denotes the aerodynamic force vector, $\mathbf{^{\text{b}}F}_{\text{grav}}\in\mathbb{R}^{\mathnormal{3}}$
denotes the force of gravity vector, $\mathbf{^{\text{b}}F}_{\text{cont}}\in\mathbb{R}^{\mathnormal{3}}$
denotes the contact force vector from ground supporting or physical
collision, and $\mathbf{^{\text{b}}F}_{\text{act},i}\in\mathbb{R}^{\mathnormal{3}}$
denotes the control force vector generated by an actuator. Noteworthy,
the above force vectors should be all transformed to the body center
and projected to the body frame.

The aerodynamic force vector $\mathbf{^{\text{b}}F}_{\text{aero}}$
is a nonlinear function determined by the relative speed of the surrounding
air $^{\text{e}}\mathbf{v}_{\text{rel}}$ as
\[
^{\text{e}}\mathbf{v}_{\text{rel}}\triangleq{}^{\text{e}}\mathbf{v}_{\text{wind}}-{}^{\text{e}}\mathbf{v}
\]
where $^{\text{e}}\mathbf{v}_{\text{wind}}$ is the wind speed from
the environment subsystem in Eq. (\ref{eq:Wind}) and $^{\text{e}}\mathbf{v}$
is the vehicle speed from the body subsystem in Eq. (\ref{eq:bodyEq}).
The high-precision aerodynamic modeling method has been well studied
in \cite{Stevens2004AirContrl}, which is compatible with all types
of vehicles such as multicopters \cite{quan2017introduction}, helicopters
\cite{cai2011unmanned} and cars \cite{rajamani2011vehicle}.

The contact force $\mathbf{^{\text{b}}F}_{\text{cont}}$ caused by
the ground supporting or physical collision can be modeled by simplifying
the vehicle body shape to a cuboid or a cylinder and simplifying the
contact surface to a spring-loaded system. By adjusting the spring
stiffness, it is convenient to simulate physical contact on objects
with different surface hardness. The terrain and obstacle information
comes from the environment subsystem output $\mathbf{y}_{\text{env}}$,
which further comes from the 3D environment subsystem $\mathbf{S}_{\text{3d}}$.

The actuator force vector $\mathbf{^{\text{b}}F}_{\text{act},i}$
can be unified described by the following nonlinear expression
\begin{equation}
\mathbf{^{\text{b}}F}_{\text{act},i}=\mathbf{f}_{\text{act},i}\left(\mathbf{\Phi}_{\text{fm}},\mathbf{y}_{\text{env}},\mathbf{y}_{\text{body}},\delta_{i}\right)\label{eq:Fact}
\end{equation}
where $\delta_{i}\in\mathbf{y}_{\text{act}}$ is the instantaneous
state of an actuator (the rotating speed of a propeller, deflection
angle of a servo system, or driving torque of a tire) from the actuator
module $\mathbf{S}_{\text{act}}$. Noteworthy, the expression of $\mathbf{f}_{\text{act},i}\left(\cdot\right)$
is also related to the vehicle state $\mathbf{y}_{\text{body}}$ and
the environment state $\mathbf{y}_{\text{env}}$, and the methods
to obtain the force model $\mathbf{f}_{\text{act},i}\left(\cdot\right)$
have been well studied in \cite{quan2017introduction,cai2011unmanned,rajamani2011vehicle}. 

\begin{figure}
\centering \includegraphics[width=0.45\textwidth]{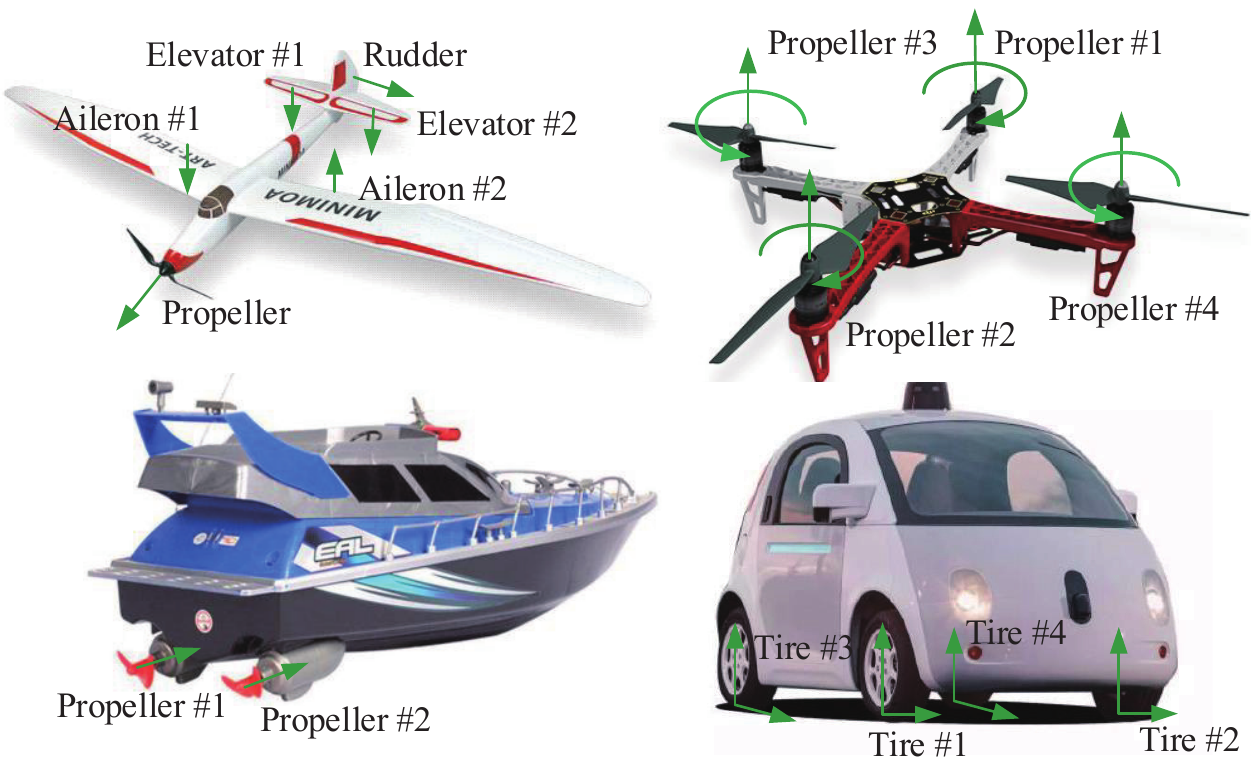}\caption{Actuator force models for different types of unmanned vehicles.}
\label{Fig05}
\end{figure}

As shown in Fig.\,\ref{Fig05}, the different types of vehicle simulation
models are mainly distinguished by the actuator types and configurations.
In a similar way, the actuator force models for different types of
vehicles in Fig.\,\ref{Fig05} can be easily obtained with proper
actuator system modeling methods.

\subsection{3D Environment Simulation Subsystem}

The 3D environment subsystem $\mathbf{S}_{\text{3d}}$ in Eq.\,(\ref{eq:SysStru})
aims to generate vision data $\mathbf{y}_{\text{3d}}$ based on the
vehicle states $\mathbf{y}_{\text{vehi}}$ from the vehicle simulation
subsystem. The vision data will be sent to the sensor subsystem to
generate data for vision sensors such as camera, radars, laser range
finder, etc.

Currently, many widely used 3D environment engines can be applied
for vehicle vision modeling. For example, the Simulink 3D Animation
Toolbox provides interfaces to conveniently access the video stream
for image process and controller design in Simulink; the Airsim \cite{Shah2018}
is developed by Microsoft$^{\circledR}$ to generate high-fidelity
visual and physical simulation environment using Epic Games$^{\circledR}$/Unreal
Engine 4 (UE4). Both Simulink 3D Animation Toolbox and Airsim can
be applied to different types of vehicles, including aircraft and
cars. There are also many 3D simulation environments exclusively developed
for specific types of vehicles. For example, the Gazebo HIL simulator
\cite{chen2018autonomous} for visual simulation of autonomous cars,
and the FlightGear \cite{Hentati2018} for aircraft simulations.

These 3D simulation engines can be applied to develop the 3D environment
subsystem to generate vision data (cameras), obstacle distance information
(rangefinders, radars), or point cloud data (Lidar). With the development
of GPU performance and 3D modeling technology, the obtained vision
data $\mathbf{y}_{\text{3d}}$ will be more and more high-fidelity
and realistic in the future, which will significantly improve the
credibility of visual simulations.

\subsection{Sensor Simulation Subsystem}

\label{subsec:Sensor-Model}

The sensor subsystem $\mathbf{S}_{\text{sens}}$ in Eq.\,(\ref{eq:SysStru})
describes the process to transform the vehicle state $\mathbf{y}_{\text{vehi}}$
and vision data $\mathbf{y}_{\text{3d}}$ to the electric signals
$\mathbf{y}_{\text{sens}}$ for the control system. Concretely, it
can also be divided into three small subsystems, including the sensor
data subsystem $\mathbf{S}_{\text{data}}$, the sensor product subsystem
$\mathbf{S}_{\text{prod}}$ and the communication subsystem $\mathbf{y}_{\text{sens}}$.
As shown in Fig.\,\ref{Fig03}, their connection relationships are
mathematically described as
\begin{equation}
\begin{array}{l}
\mathbf{y}_{\text{data}}=\mathbf{S}_{\text{data}}\left(\mathbf{u}_{\text{data}}\right),\,\mathbf{u}_{\text{data}}=\mathbf{u}_{\text{sens}}=\left\{ \mathbf{y}_{\text{vehi}},\mathbf{y}_{\text{3d}}\right\} \\
\mathbf{y}_{\text{prod}}=\mathbf{S}_{\text{prod}}\left(\mathbf{u}_{\text{prod}}\right),\,\mathbf{u}_{\text{prod}}=\mathbf{y}_{\text{data}}\\
\mathbf{y}_{\text{\text{sens}}}=\mathbf{y}_{\text{com}}=\mathbf{S}_{\text{com}}\left(\mathbf{u}_{\text{\text{com}}}\right),\,\mathbf{u}_{\text{\text{com}}}=\mathbf{y}_{\text{prod}}
\end{array}\label{eq:sensModel}
\end{equation}
where $\mathbf{y}_{\text{data}}$ contains ideal data for sensors
(e.g., the acceleration of accelerometers, the magnetic field of magnetometers,
and the image or point cloud data of vision sensors), and $\mathbf{y}_{\text{prod}}$
is the sensor signals after adding detailed product features (e.g.,
noise level, temperature drift, failure mode, and camera distortion),
and $\mathbf{y}_{\text{sens}}$ is the binary electrical signals transmitted
to the control system for position and attitude estimation.

\subsubsection{Sensor Data Subsystem}

As described in Eq.\,(\ref{eq:sensModel}), the sensor data subsystem
$\mathbf{S}_{\text{data}}$ generates the sensor data $\mathbf{y}_{\text{data}}$
(applicable for a class of sensor products) based on the vehicle state
$\mathbf{y}_{\text{vehi}}$ and vision data $\mathbf{y}_{\text{3d}}$.
Transformations usually required to obtained the sensor data from
the vehicle states and vision data. For example, accelerometers measure
the specific force (the difference between the acceleration of the
aircraft and the gravitational acceleration) \cite[p. 122]{beard2012small}
instead of the vehicle acceleration $\mathbf{^{\text{b}}\dot{v}}\in\mathbf{y}_{\text{vehi}}$.
Similar computation expressions are applied to other types of sensors,
such as the GPS Modules (longitude and latitude obtained from the
vehicle position), electronic compasses (the magnetic field intensity
obtained from the attitude and global position of the vehicle), optical
flow sensors (relative velocity obtained from image stream), etc.

\subsubsection{Sensor Product Subsystem}

The sensor product subsystem $\mathbf{S}_{\text{prod}}$ is developed
to add product features (e.g., noise, vibration, and calibration)
to the sensor data $\mathbf{y}_{\text{data}}$ obtained from the above
sensor data subsystem $\mathbf{S}_{\text{data}}$. Given the same
sensor data, different sensor products may obtain different results
due to product features, so the senor product subsystem is necessary. 

In most cases, given the ideal sensor data $\mathbf{x}_{\text{m}}\in\mathbf{y}_{\text{data}}$,
the noise feature is mainly reflected in the noise $\mathbf{n}_{\text{a}}$
and bias drift $\mathbf{b}_{\text{a}}$ of the measured value $\mathbf{x}_{\text{m}}^{\prime}$,
which can be described as \cite[p. 151]{quan2017introduction}
\begin{equation}
\begin{cases}
\mathbf{x}_{\text{m}}^{\prime} & =\mathbf{x}_{\text{m}}+\mathbf{b}_{\text{a}}+\mathbf{n}_{\text{a}}\\
\mathbf{\dot{b}}{}_{\text{a}} & =\mathbf{n}_{\text{b}}
\end{cases}\label{eq:nose}
\end{equation}
where $\mathbf{n}_{\text{a}}\sim N\left(0,\sigma_{\text{a}}^{2}\right)$
and $\mathbf{n}_{\text{b}}\sim N\left(0,\sigma_{\text{b}}^{2}\right)$
are zero-mean Gaussian noise vectors for inertial sensors. The standard
deviation parameters $\sigma_{\text{a}},\sigma_{\text{b}}$ can be
found in the datasheet document of a sensor product or obtained by
system identification with the actual sensor output signals. 

If the system is not affected by vibrations, $\sigma_{\text{a}}$
can be modeled as a constant value. When the vibration feature of
a sensor is considered, the measuring noise $\mathbf{n}_{\text{a}}$
may also be affected by the vibration from may sources (e.g., engines,
motors, and fuselage).Therefore, the standard deviation $\sigma_{\text{a}}$
is not always constant value, which should be modeled based on the
actual system characteristics.

The calibration feature is mainly determined by the working environment
of the installation configuration a sensor, which can be described
as \cite[p. 149]{quan2017introduction}
\begin{equation}
\mathbf{x}_{\text{m}}^{\prime\prime}=\mathbf{T}_{\text{e}}\mathbf{K}_{\text{e}}\left(\mathbf{x}_{\text{m}}^{\prime}+\mathbf{p}_{\text{e}}\right)\label{eq:eror}
\end{equation}
where $\mathbf{p}_{\text{e}}$ is a constant vector for the position
installation deviation, $\mathbf{T}_{\text{e}}$ is a rotation matrix
for the installation deviation, $\mathbf{K}_{\text{e}}$ is a diagonal
matrix for the scale deviation, and $\mathbf{x}_{\text{m}}^{\prime\prime}\in\mathbf{y}_{\text{prod}}$
is the final output data of a sensor. Eqs.\,(\ref{eq:nose})(\ref{eq:eror})
are applicable to most types of sensors to simulate the properties
of real sensor products.

There are also many methods to add product features for vision sensors.
For example, the methods to add camera features (e.g., blurs, distortions,
and noises) to an ideal image is introduced in \cite{Kucis2012}.
Other environment factors (e.g., lighting, reflection, and fogging)
can be simulated by the latest 3D engines, such as UE4. 

\subsubsection{Communication Subsystem}

The communication subsystem is developed to transform the sensor data
with product features $\mathbf{y}_{\text{prod}}$ to binary electronic
signals $\mathbf{y}_{\text{\text{sens}}}$ for the control system.
The outputs $\mathbf{y}_{\text{prod}}$ of the above sensor product
subsystem $\mathbf{S}_{\text{prod}}$ are decimal numerical signals,
but binary electronic signals are required for the communication requirements
between the control system and other hardware. There are many communication
interfaces and protocols widely used in the vehicle control systems,
such as SPI, Inter-Integrated Circuit (I$^{2}$C) \cite{leens2009introduction},
Controller Area Network (CAN), Universal Asynchronous Receiver-Transmitter
(UART), and Pulse-Width Modulation (PWM).

\begin{figure}
\centering \includegraphics[width=0.45\textwidth]{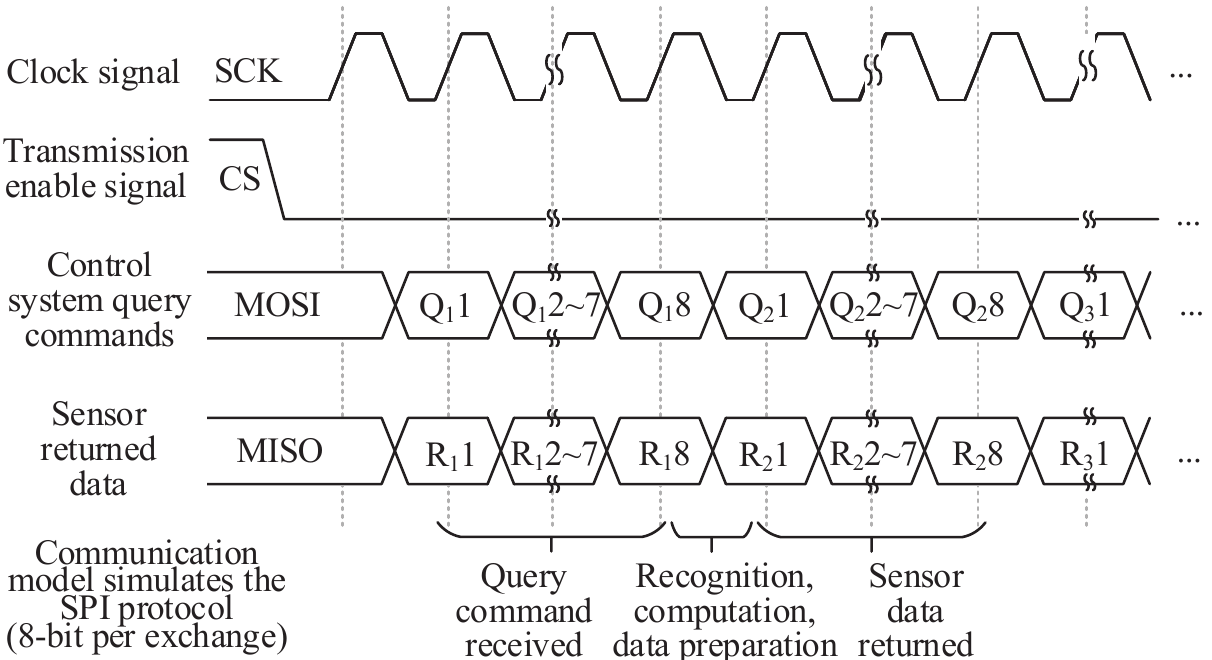}\caption{Communication subsystem model for SPI buses.}
\label{Fig07}
\end{figure}

The mathematical model for a communication interface is usually simple,
but it is hard to be simulated by a CPU-based simulation computer
due to the extremely high real-time update frequency and bandwidth
requirements. Taking the SPI interface as an example, the SPI interface
uses four signal wires to exchange information between the master
device (the main processor of the control system) and the slave device
(onboard sensors). As shown in Fig.\,\ref{Fig07}, the sensor has
to finish the command recognition, measured data computation, and
output data preparation within a small interval (after the previous
byte data received and before the following byte data to be sent).
In a real sensor chip, the above process is instantaneously realized
by analog circuits whose time consumption can be treated as infinitely
small. However, for a real-time simulation system, it usually requires
a real-time update frequency at the nanosecond level to ensure that
the sensor signals are correctly computed, prepared, and transmitted.
For the performance requirements, this paper uses FPGA system to simulate
all sensor communication features (e.g., data transmission, chip recognition,
programmable setting functions), which ensures all the sensor hardware
related low-level test cases can be simulated by the simulation platform.

\subsection{Model Identification and Validation}

The mathematical model of a subsystem can be classified into three
types: the static models, the dynamic models, and the statistical
models. (i) The static models are described by constant parameters
that can usually be precisely measured by professional equipment.
These parameters include the mass, moment of inertia, air density,
aerodynamic coefficients, etc. (ii) The dynamic models are described
by dynamic equations that are usually difficult to be measured by
equipment directly. System identification methods \cite{tischler2018system,remple2006aircraft}
are widely used in obtaining dynamic parameters in aircraft and vehicle
systems. (iii) The statistical models are described by statistical
values or stochastic processes (e.g., the sensor noise, and the atmospheric
turbulence). The statistical models can be obtained by spectral analysis
or statistical analysis of the output signals with enough sample data.
For example, the expectation and variance of the White Gaussian noise
of a sensor product in Eq.\,(\ref{eq:nose})(\ref{eq:eror}) can
be obtained by its output signals \cite{el2008analysis}.

In practice, by using advanced modeling methods with enough experimental
data, any vehicle system can be modeled arbitrarily close to the real
system. However, a high-fidelity mathematical model usually indicates
more computational complexity and cost (higher dimension and nonlinearity),
so abstraction and simplification methods are important for simulation
systems. The simulation credibility is the most important index in
the trade-off between precision and complexity. In our previous research
\cite{Dai2019Simulation}, by feeding the same signals to both the
simulation system and the real system and analyzing the simulation
errors, a simulation credibility assessment method is proposed to
assess the simulation credibility of a simulation system from multiple
aspects, such as system performance, time-domain response, and frequency-domain
response. A normalized assessment index is proposed to score the simulation
credibility of a subsystem under a uniform assessment criterion. The
proposed method is efficient to verify and validate the simulation
effect of each subsystem and the whole simulation system, which is
used in this paper to ensure the credibility of the obtained vehicle
subsystem models.

\section{\textit{Real-time HIL Test Platform with MBD}}

\label{sec:3}This section presents the hardware structure and development
framework of the proposed HIL test platform in Fig. \ref{Fig03} successively.

\subsection{Hardware Structure of the HIL Platform}

The platform hardware consists of three parts.

\subsubsection{\textit{Real-time} Simulation Computer}

A real-time simulation computer (also called real-time simulator)
is a special type of computer, running Real-Time Operation System
(RTOS) to ensure the simulation systems run at the same speed as the
actual physical system. The latest COTS simulation computers usually
provide a CPU-based system and an FPGA-based system for different
requirements of real-time update frequency; the CPU-based system is
better at running complex simulation models with moderate frequency
requirements (usually smaller than 100KHz); the FPGA-based system
is better at running simple simulation models with extremely high
frequency (usually larger than 100MHz). By combining the advantages
of the above two systems, the hardware structure of the proposed test
platform is presented in Fig.\,\ref{Fig03}, where the CPU-based
system is applied to run the vehicle simulation subsystem presented
in Section \ref{subsec:Vehicle-Simulation-Subsystem} and the FPGA-based
system is applied to run the sensor simulation subsystem presented
in \ref{subsec:Sensor-Model}.

\subsubsection{Control System}

The control system is the test object of the proposed test platform.
The control system computes control signals for driving the actuators
according to the vehicle states measured and estimated by different
sensors. To ensure the control system can work normally in the proposed
test platform, the original sensors should be blocked (or removed),
and the sensor pins on the control system should be reconnected to
the FPGA-based system of the real-time simulation computer. Then,
the control system can receive the simulated sensor chip signals for
full-function operation. The above process only needs to know the
brands and models of sensors used in the control system, and has no
requirement to access the source code or internal hardware structure.
Therefore, it is practical to perform black-box testing for different
control system products from different unmanned vehicle companies.

\subsubsection{Host Computer}

A high-fidelity 3D simulation environment is also essential for training
or testing the top-level algorithms, including computer vision, machine
intelligence, and decision making systems. Therefore, a host computer
with high-performance Graphics  Processing Units (GPUs) and realistic
3D engines are used in this paper to generate vision signals to the
real-time simulation computer. The latest high-end consumer GPUs (such
as NVIDIA GTX2080) have been powerful enough to generate high-resolution
(larger than 4K) video streaming with update frequency more than 100Hz,
which is capable of simulating most vision sensors. Besides, the host
computer also takes responsibility for running other auxiliary programs
such as model parameter configuration program, 3D display program,
ground control program, etc. Noteworthy, for different data bandwidth
and real-time requirements, the connection and communication among
the simulation computer and the host computer can be realized by optical
fibers, network cables, serial ports, etc.

\subsection{Development Framework with MBD}

\label{subsec:3.2}

\subsubsection{Modular Programming}

In practice, developing a complex simulation software through hand-coding
is a difficult and unreliable task due to too many mathematical operations.
Any coding mistake, logical mistake, or unknown vulnerability may
lead to wrong or inaccurate simulation results. The types and amounts
of unmanned vehicles will be far more than manned vehicles, and the
development cycles are required to be much shorter, so the traditional
hand-coding methods are no longer suitable for developing simulation
software for unmanned vehicles. As a result, modular (also described
as graphical or visual) programming methods have been widely used
in many MBD tools (e.g., MathWorks$^{\circledR}$/Simulink and NI$^{\circledR}$/LabVIEW).
The whole simulation system presented in Section \ref{sec:2} has
been divided into many simple and independent subsystems, and each
subsystem can be easily realized by a visual module or block in the
above MBD tools, which makes it easy to develop and test a complicated
vehicle simulation system. 

\subsubsection{Automatic Code Generation}

Modular programming and automatic code generation are two of the most
significant features of MBD tools such as Simulink, LabVIEW, and UE4.
Simulink is better at developing complex simulation systems (such
as unmanned vehicle simulation systems), and LabVIEW is better at
developing hardware-closed simulation systems (such as sensors, circuits
and communication signals) and deploying the simulation program to
the real-time simulator. For example, the Aerospace Blockset in Simulink
\cite{gage2018nasa} provides many demos for quickly developing vehicle
simulation systems, such as aircraft and multicopter. There are many
3D engines for developing high-fidelity vehicle 3D simulation environments.
Among these engines, UE4 provides modular visual programming functions,
so the UE4 environment is selected to develop 3D simulation scenes
for the HIL platform.

\begin{figure}
\centering \includegraphics[width=0.45\textwidth]{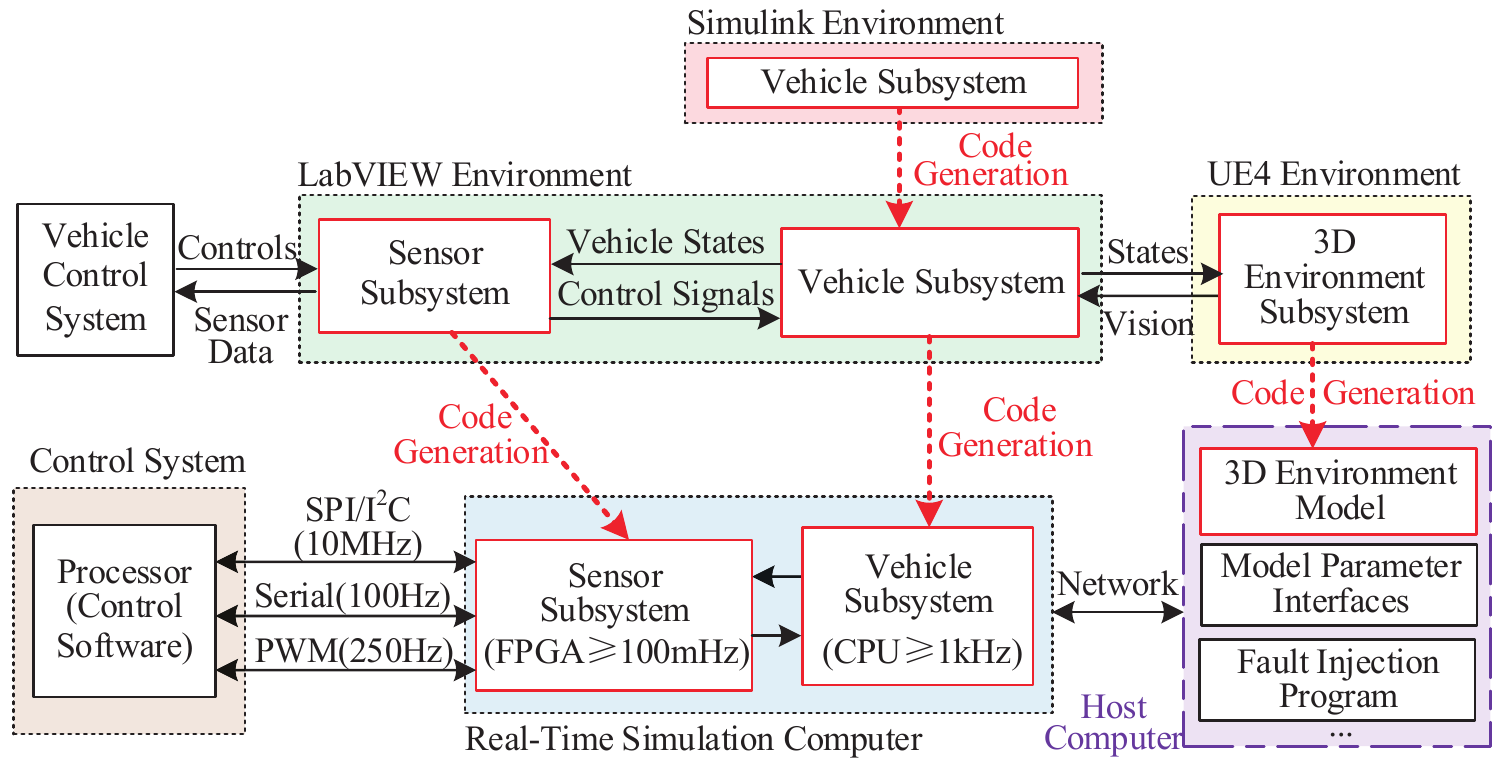}\caption{Code generation and deployment framework for simulation software of
HIL testing platforms.}
\label{Fig09}
\end{figure}

In order to take full advantages of both MBD tools, the development
process for simulation system software is shown in Fig.\,\ref{Fig09}.
The development process is divided into the following steps: (i) developing
and verifying the vehicle simulation model in Simulink Environment;
there are many powerful verifying tools in Simulink such as requirements
traceability, code coverage check, document generation, etc., which
guarantee the simulation software meets the standards and guidelines
such as DO-178C; (ii) compiling the vehicle simulation subsystem into
code, and importing it to the LabVIEW environment; (iii) building
the sensor subsystem in LabVIEW and building the interfaces to communicate
with other systems; (iv) generating code and executable files to deploy
them to the real-time simulation computer; (v) developing the 3D simulation
environment in UE4 and deploy it to the host computer. The whole development
process in Fig.\,\ref{Fig09} is efficient and reliable because all
coding and deploying operations are automatically finished by MBD
tools without much human intervention. Therefore, it is convenient
to replace some models and rebuild the simulation system for different
types of vehicles or control systems.

\section{Verification and Application}

\label{sec:4}In this section, a real-time HIL test platform is first
developed for multicopter UAVs based on the proposed test platform.
Then, the simulation accuracy is verified through a series of experiments.
In the end, several successful applications are presented to verify
the feasibility and practicability of the proposed methods in this
paper.

Four videos have been published to demonstrate the development, experimental
verification, and application process in this section for the proposed
platform. The first video gives an overall introduction of the proposed
platform along with several applications on different multicopter
control systems:\vspace{-0.15cm}

\begin{center}
\uline{\url{https://youtu.be/nXAoLdPzz_I}}\vspace{-0.15cm}
\par\end{center}

\noindent The second video presents several demos of applying the
proposed HIL platform to different types of vehicles (cars, copters,
aircraft) with multi-vehicle traffic environment simulation:\vspace{-0.15cm}

\begin{center}
\uline{\url{https://youtu.be/xQUnkqH29qU}}\vspace{-0.15cm}
\par\end{center}

\noindent The third video presents the automatic fault injection,
safety testing, and safety assessment demos:\vspace{-0.15cm}

\begin{center}
\uline{\url{https://youtu.be/MHieyE3hbHY}}\vspace{-0.15cm}
\par\end{center}

\noindent The fourth video introduces the MBD modeling and development
process of the platform with experiments to quantitatively verify
its simulation credibility:\vspace{-0.15cm}

\begin{center}
\uline{\url{https://youtu.be/ChNtkb5rrQs}}\vspace{-0.15cm}
\par\end{center}

\noindent We have also published the MATLAB source code and the detailed
modeling tutorial to Github: \vspace{-0.15cm}

\begin{center}
\uline{\url{https://github.com/XunhuaDai/CopterSim}}\vspace{-0.15cm}
\par\end{center}

\noindent Readers can use it to rapidly develop SIL or HIL simulation
systems for different types of unmanned vehicles by modifying the
aerodynamic model and the actuator model.

\begin{figure}
\centering \includegraphics[width=0.45\textwidth]{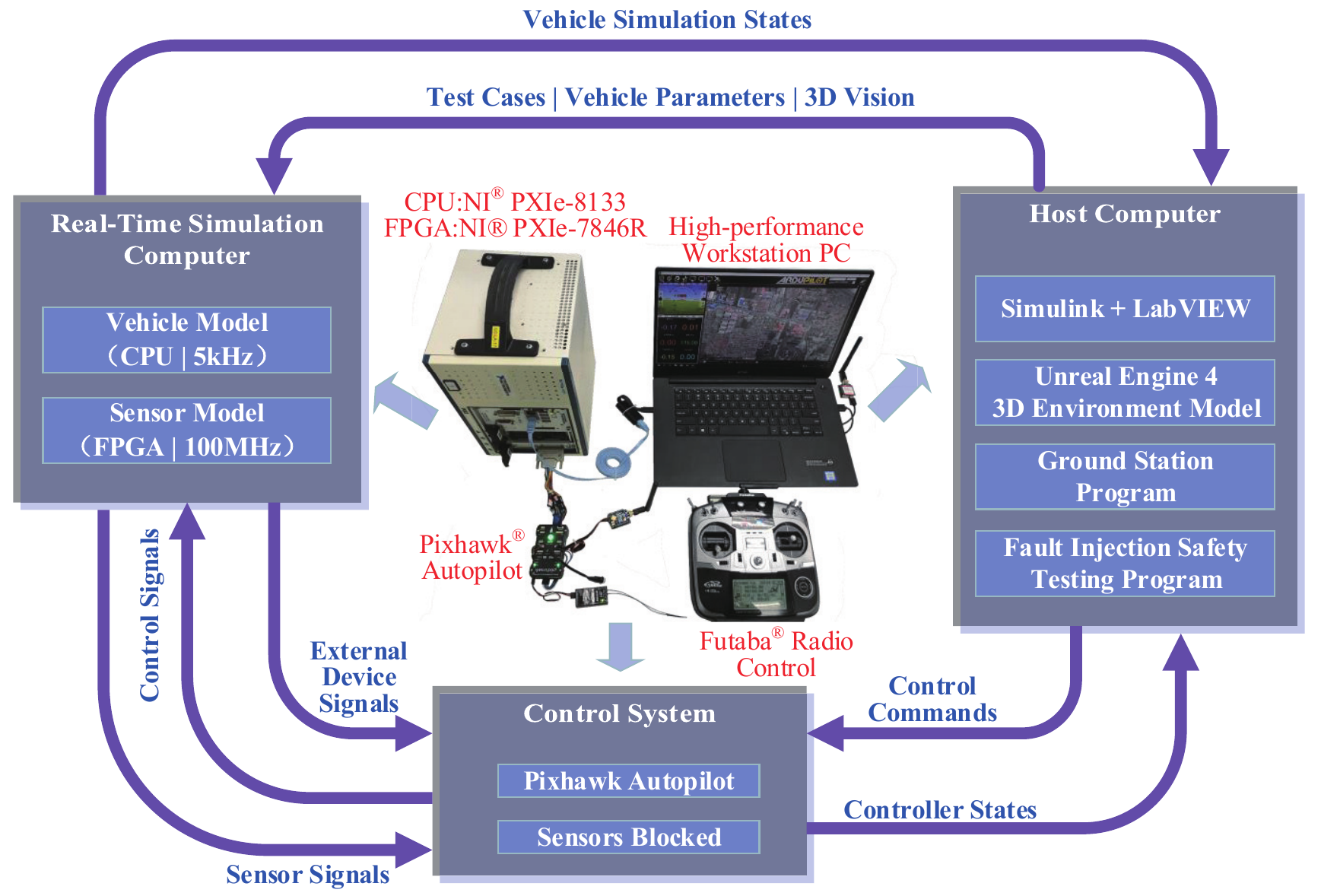}\caption{Hardware composition of the real-time HIL test platform}
\label{Fig12}
\end{figure}

\subsection{Platform Implementation}

\subsubsection{Hardware Composition}

Based on the hardware structure presented in Fig.\,\ref{Fig03},
a real-time HIL test platform is developed as shown in Fig.\,\ref{Fig12}.
The real-time simulation computer adopted in the platform is the NI$^{\circledR}$/PXI
simulator with CPU board: PXIe-8133 (Intel Core I7 Processor, PharLap
ETS Real-Time System) and FPGA I/O Module: PXIe-7846R. The host computer
is a high-performance workstation PC with professional GPU. The control
system is Pixhawk autopilot, which is one of the most popular open-source
autopilot hardware systems for small unmanned vehicles. All the onboard
sensors (IMU, magnetometer, barometer, etc.) and external sensors
(GPS, rangefinder, camera, etc.) of the Pixhawk hardware are blocked,
and the sensor pins are reconnected to the FPGA I/Os to receive the
simulated sensor signals through interfaces including SPI, PWM, I$^{2}$C,
UART, UBX, etc. In the real-time simulation computer, the update frequency
of the vehicle simulation model is up to 5 kHz, and the update frequency
of sensor simulation model is up to 100 MHz, whose performance is
fast enough for HIL simulations of most commercial control systems.
The communication among between the host computer and the real-time
simulation computer is realized by network cables with TCP and UDP
protocols. 

\subsubsection{Software Development}

\begin{figure}
\centering \includegraphics[width=0.42\textwidth]{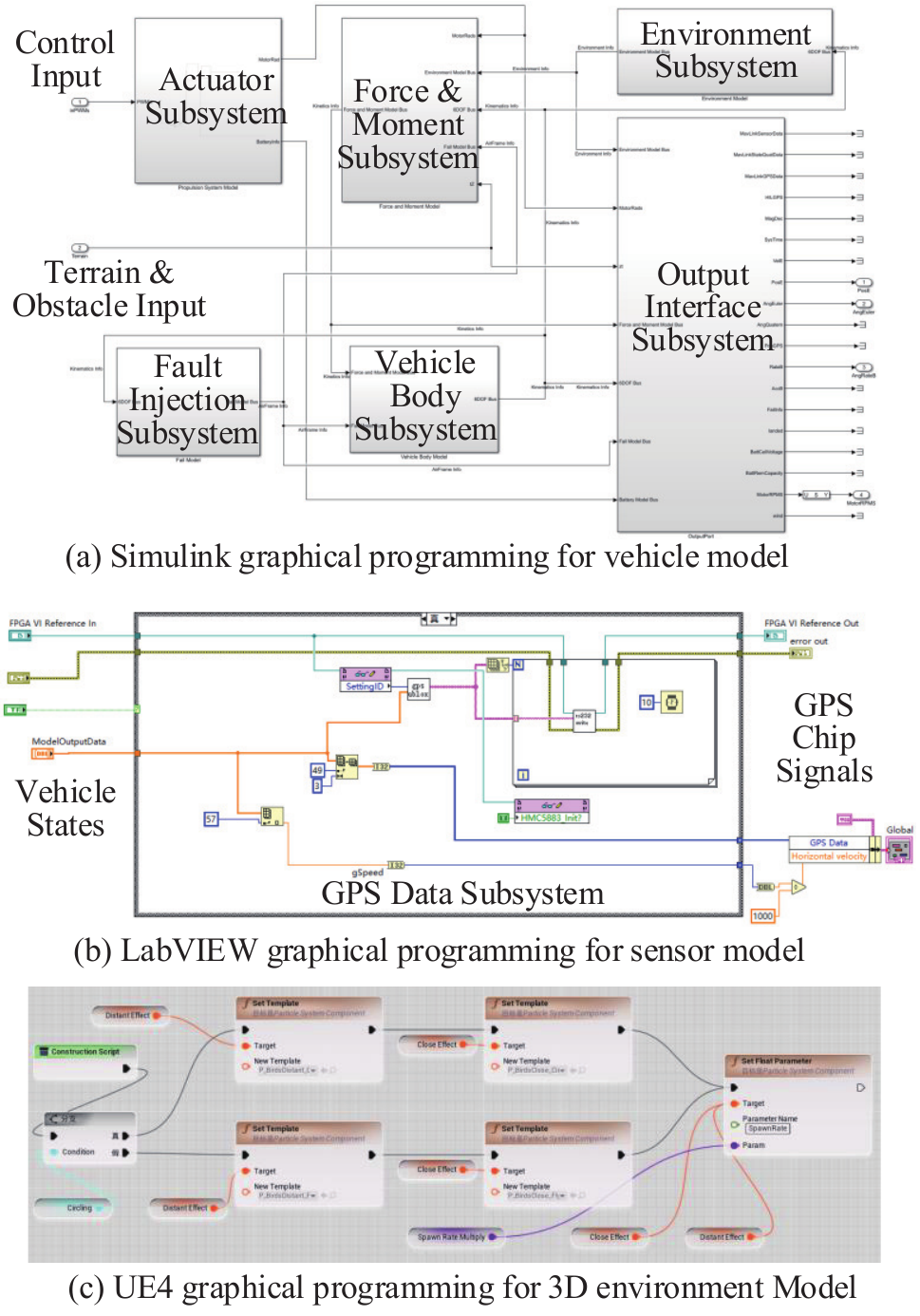}\caption{Modular visual programming environments in Simulink, LabVIEW, and
UE4.}
\label{Fig13}
\end{figure}

The simulation software of the real-time HIL test platform is developed
based on the MBD method in Section \ref{subsec:3.2}. The Simulink
is selected to develop the vehicle simulation model because it is
the most professional and widely used software for vehicle dynamic
system development; the LabVIEW is selected to develop the sensor
simulation model because it is efficient and convenient in real-time
simulation system development; the UE4 (version: 4.22) is selected
to develop the 3D environment model because it is one of the most
realistic 3D engines in the development of simulation systems, games
and VR systems. The Simulink, LabVIEW, and UE4 developing environments
all provide convenient modular visual programming environments (see
Fig.\,\ref{Fig13}) and automatic code generation technology for
the development of simulation systems, which are perfect for the implementation
of the model-based design method. After the simulation models are
all developed, the code generation and deployment framework in Fig.\,\ref{Fig09}
is applied to deploy the simulation software to the real-time HIL
test platform presented in Fig.\,\ref{Fig12}.

\subsubsection{High-fidelity 3D Simulation Environment}

\begin{figure}
\centering \includegraphics[width=0.5\textwidth]{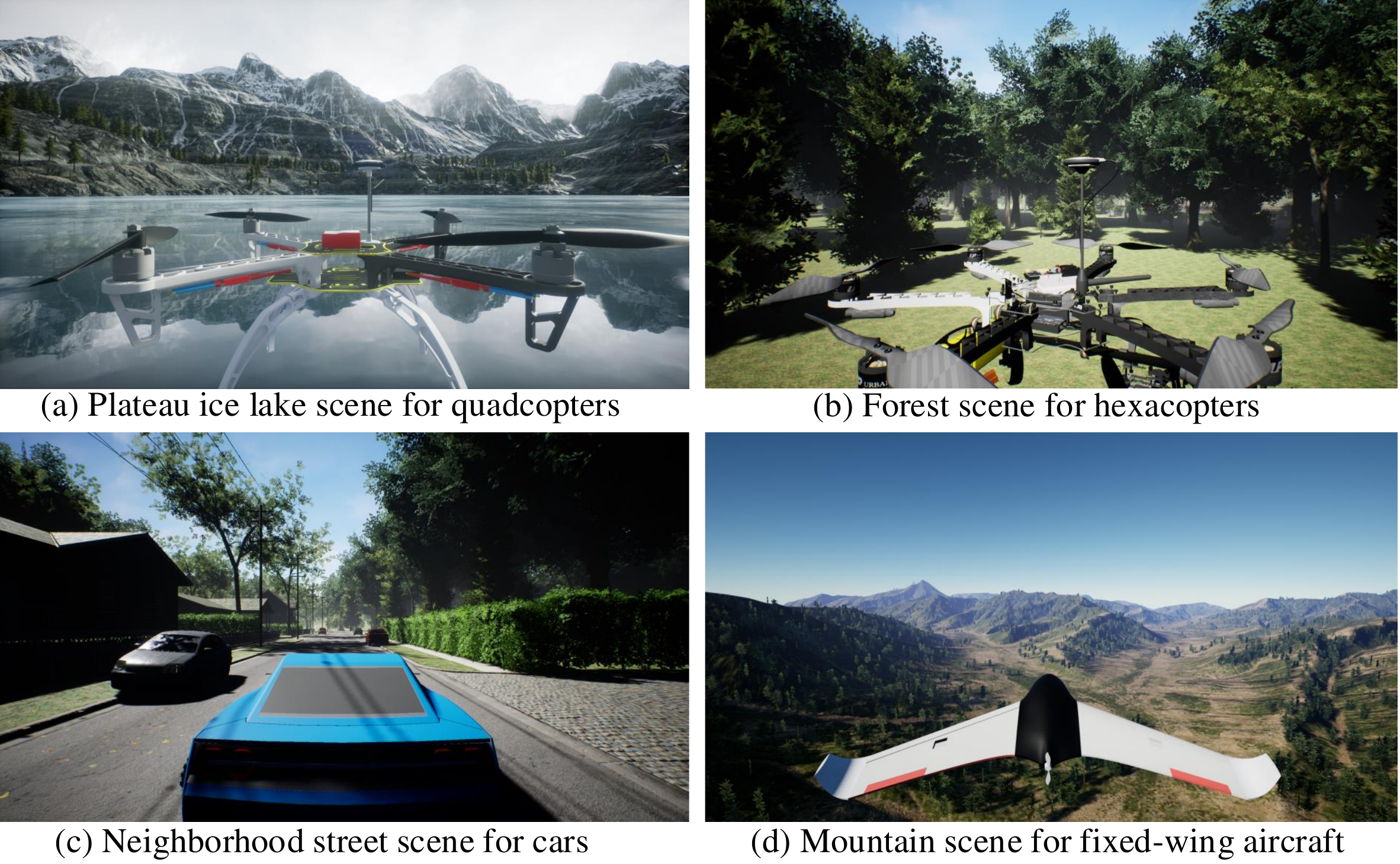}\caption{3D simulation scenes developed by UE4 for different types of unmanned
vehicles.}
\label{Fig16}
\end{figure}

The fidelity of the 3D simulation model is important for testing the
vision-related functions of control systems, such as visual data processing,
obstacle avoidance, safety decision-making, etc. With UE4, it is easy
to develop high-fidelity 3D scenes for different types of vehicle
in different environments. For example, as shown in Fig.\,\ref{Fig16},
we have developed several 3D simulation scenes in UE4 for the HIL
test platform. According to the comparisons with experiments, the
display effect presented in Fig.\,\ref{Fig16} has been realistic
enough for most unmanned vehicle systems to simulate the real indoor
or outdoor scenes.

\subsection{Experiments and Verification}

\subsubsection{Platform Feature Verification}

The Pixhawk autopilot supports for running different types of flight
control software systems (e.g., PX4, Ardupilot, and other embedded
control software systems) in it, and also supports for controlling
different types of unmanned vehicles (e.g., multicopters, small cars,
and fixed-wing aircraft). Besides, the Pixhawk autopilot has a series
of available hardware configurations for certain performance requirements,
such as Pixhawk 1, Pixhawk 4, etc. To verify the proposed modeling
method and test platform, we first apply the proposed unified modeling
method in Section \ref{sec:2} to develop the simulation models for
different types of unmanned vehicles, including multicopters, small
cars, and fixed-wing aircraft. Then, these simulation models are deployed
to the HIL test platform with the MBD framework presented in Section
\ref{sec:3}. Finally, a series of tests are performed for the Pixhawk
systems with various combinations of hardware configurations, software
systems, and vehicle types. In our experimental tests, the four advantages
of the proposed method (including extensibility, comprehensiveness,
verification, and standardization) concluded in Section \ref{subsec:3.2}
are verified with the proposed modeling method and simulation test
platform. 

\subsubsection{Experimental Setup}

For simplicity, a quadcopter control system will be selected as the
representative tested object in this section to perform quantitative
verification for the proposed methods. The control system is the most
widely used open-source autopilot system for small-scale unmanned
vehicles, and the detailed configuration is Pixhawk 1 hardware (MCU:
STM32F427, sensor: MPU6000, MS5611, LSM303D, L3GD20H, Ublox-M8N, etc.)
with the PX4 control software. The quadcopter UAV is selected because
it is the most representative unmanned vehicle type that covers the
model characteristics (e.g., aerodynamics, ground collision, kinematics,
and dynamics) and operating environments (e.g., near-ground, mid-air,
indoor, outdoor, hovering flight, and forward flight) of most unmanned
vehicles.

\begin{figure}
\centering \includegraphics[width=0.45\textwidth]{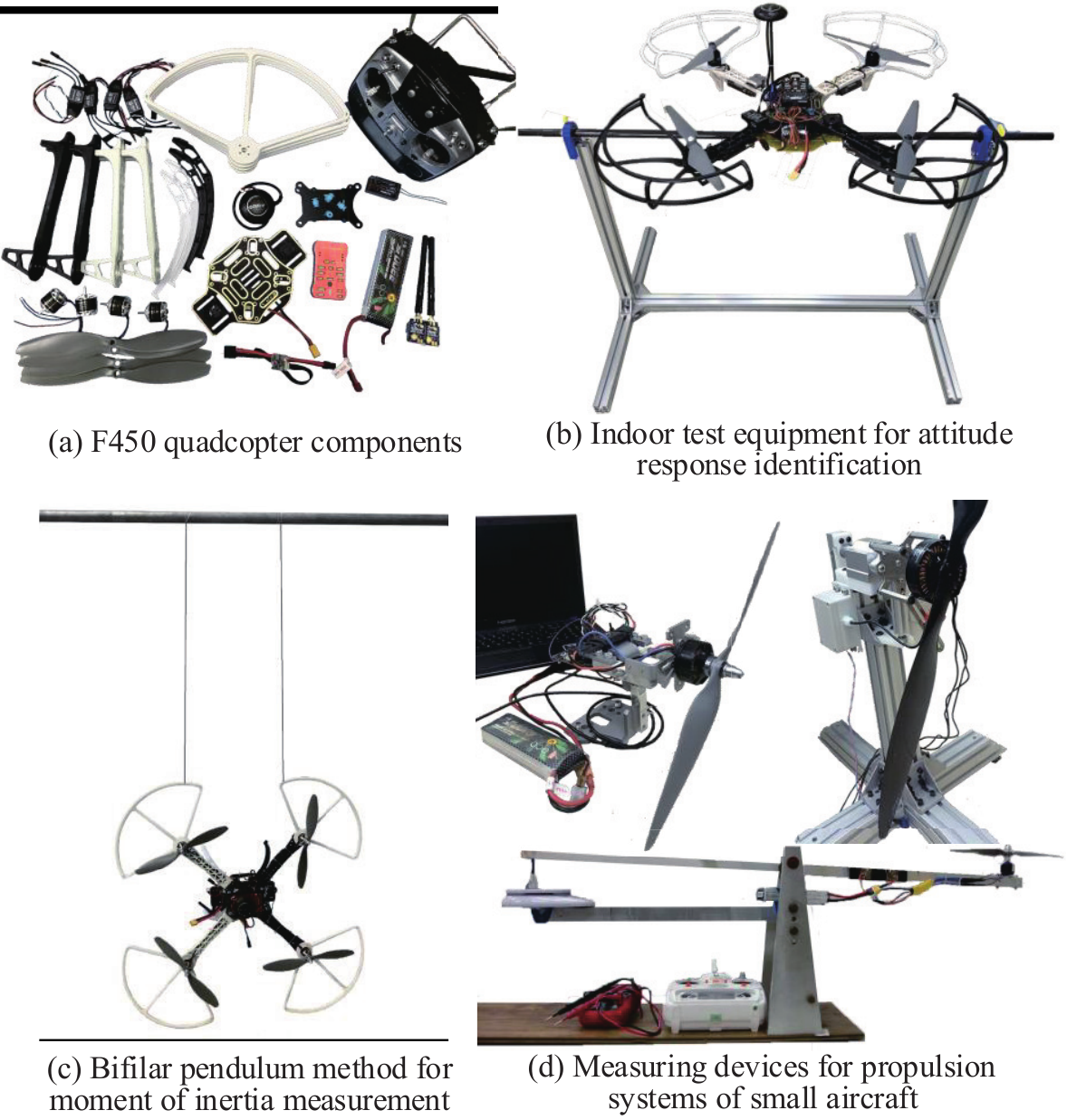}\caption{Verification equipment for the proposed HIL test platform.}
\label{Fig14}
\end{figure}

The experimental setup is presented in Fig.\,\ref{Fig14}, where
an F450 quadcopter airframe (diagonal length: 450mm, weight: 1.4kg,
propulsion system: DJI E310, battery: LiPo 3S 4000mAh) is selected
as the experimental subject whose component diagram is shown in Fig.\,\ref{Fig14}(a).
In order to test the attitude dynamics and aerodynamics of the quadcopter,
an indoor test bench (see Fig.\,\ref{Fig14}b) is developed with
the quadcopter fixed to a stiff stick (through the center of mass)
with smooth bearings to minimize friction. The quadcopter is free
to smoothly rotating along one axis, which makes it possible to perform
sweep-frequency testing for system identification and uniform rotation
testing for roll damping coefficient measurement. Figs.\,(c)(d) present
the test benches to measure the moment of inertia and the propulsion
system parameters of small-scale unmanned vehicles, and the detailed
measuring methods can be found in our previous work \cite[pp. 121-143]{quan2017introduction}.
Besides, lots of outdoor flight tests are also performed to obtain
the aerodynamic coefficients of the tested quadcopter and verify the
platform simulation results with actual flight results. The experimental
process and results have also been presented in the video mentioned
in the last subsection. 

\subsubsection{Simulation validation}

\begin{figure}
\centering \includegraphics[width=0.48\textwidth]{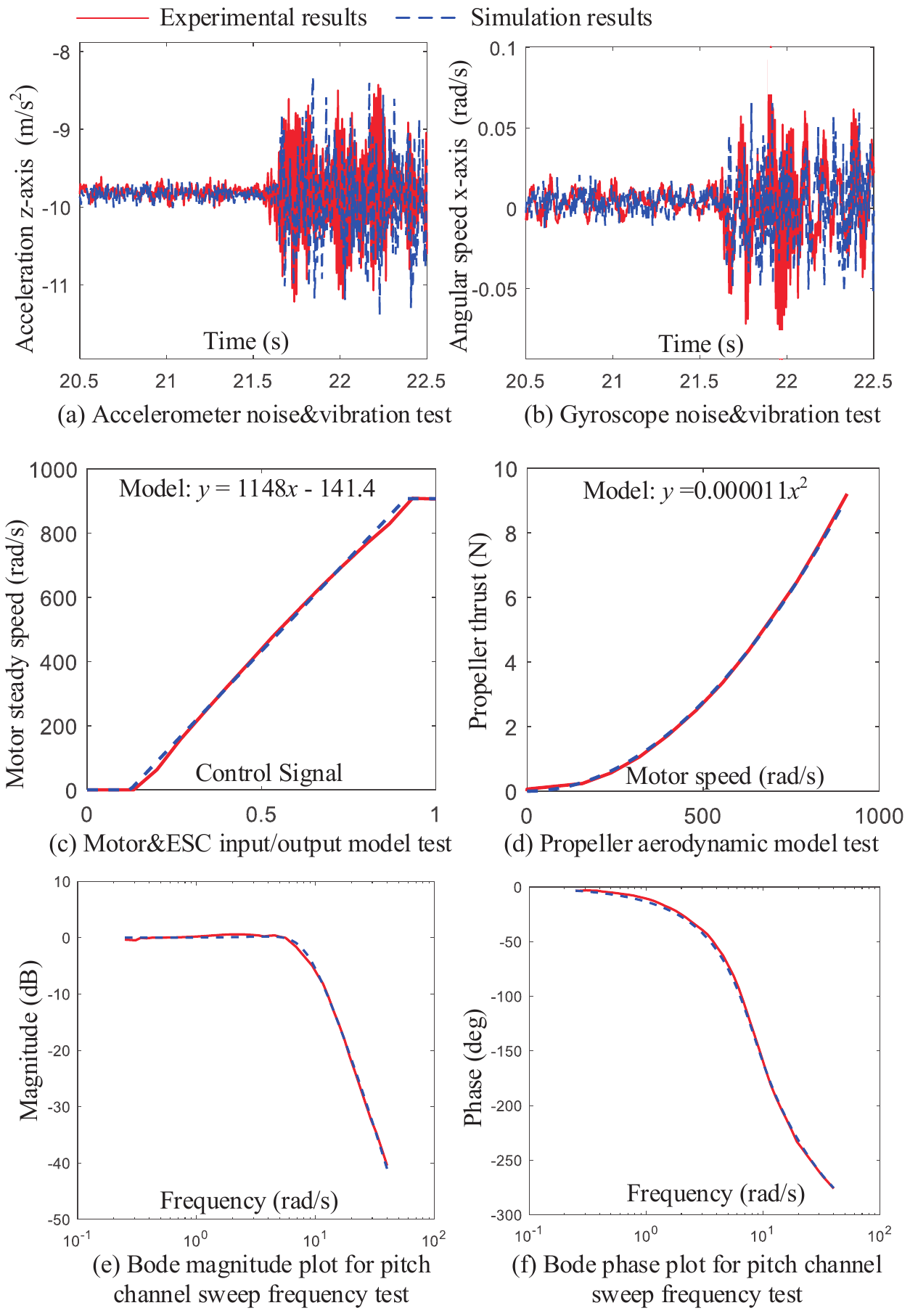}\caption{Comparison experiments to verify the simulation accuracy of the HIL
test platform.}
\label{Fig16-1}
\end{figure}

A series of comparative experiments are performed with the experimental
test benches (see Fig.\,\ref{Fig14}) and the HIL simulation test
platform (see Fig.\,\ref{Fig12}) to verify proposed modeling method
and test platform, and several representative comparison results are
presented in Fig.\,\ref{Fig16-1}. In Figs.\,\ref{Fig16-1}(a)(b),
the accelerometer and gyroscope data obtained from the IMU sensor
(MPU 6000 in Pixhawk) are presented to verify the sensor modeling
methods in Section \ref{subsec:Sensor-Model}. In the simulation and
the experiment, the motor throttle is stepped to 50\% at 21.5s to
test the vibration effect caused by the high-speed rotating motors
and propellers. In Figs.\,\ref{Fig16-1}(c)(d), the propeller actuator
modeling method in Section \ref{subsec:Vehicle-Simulation-Subsystem}
is verified with the propulsion system test benches in Fig.\,\ref{Fig14}(d).
Besides, sweep frequency tests are performed for the quadcopter pitch
channel with the test bench presented in Fig.\,\ref{Fig14}(c), and
the magnitude and phase curves of the Bode plots obtained by the CIFER$^{\circledR}$
software \cite{remple2006aircraft} are presented in Figs.\,\ref{Fig16-1}(e)(f),
respectively. From the comparative results in Fig.\,\ref{Fig16-1},
the following conclusions can be obtained. (i) The FPGA-based HIL
test platform allows using the same control system in both experiments
and simulations, which minimizes the disturbing factors for result
analysis and makes it convenient to acquire control system data for
comparison and assessment. (ii) From the perspective of qualitative
analysis, the obtained HIL test results are close to the experimental
results in Fig.\,\ref{Fig16-1} from both time-domain and frequency-domain
aspects, which verify the simulation accuracy and credibility of the
proposed modeling method and test platform. (iii) The quantitative
simulation credibility assessment method proposed in our previous
work \cite{Dai2019Simulation} is applied in this paper to assess
and improve the simulation credibility of the HIL platform, which
ensures that a high matching degree (the credibility index in \cite{Dai2019Simulation})
larger than 90\% (where 60\% presents the minimum accuracy requirement,
and 100\% presents a perfect match) is obtained by analyzing the results
between the test platform and real experimental system from the quantitative
perspective.

\subsection{Method Applications}

This subsection presents several successful applications with the
proposed test platform to increase the development, testing, and validation
efficiency of multicopters.

\subsubsection{Rapid Prototyping}

\begin{figure}
\centering \includegraphics[width=0.45\textwidth]{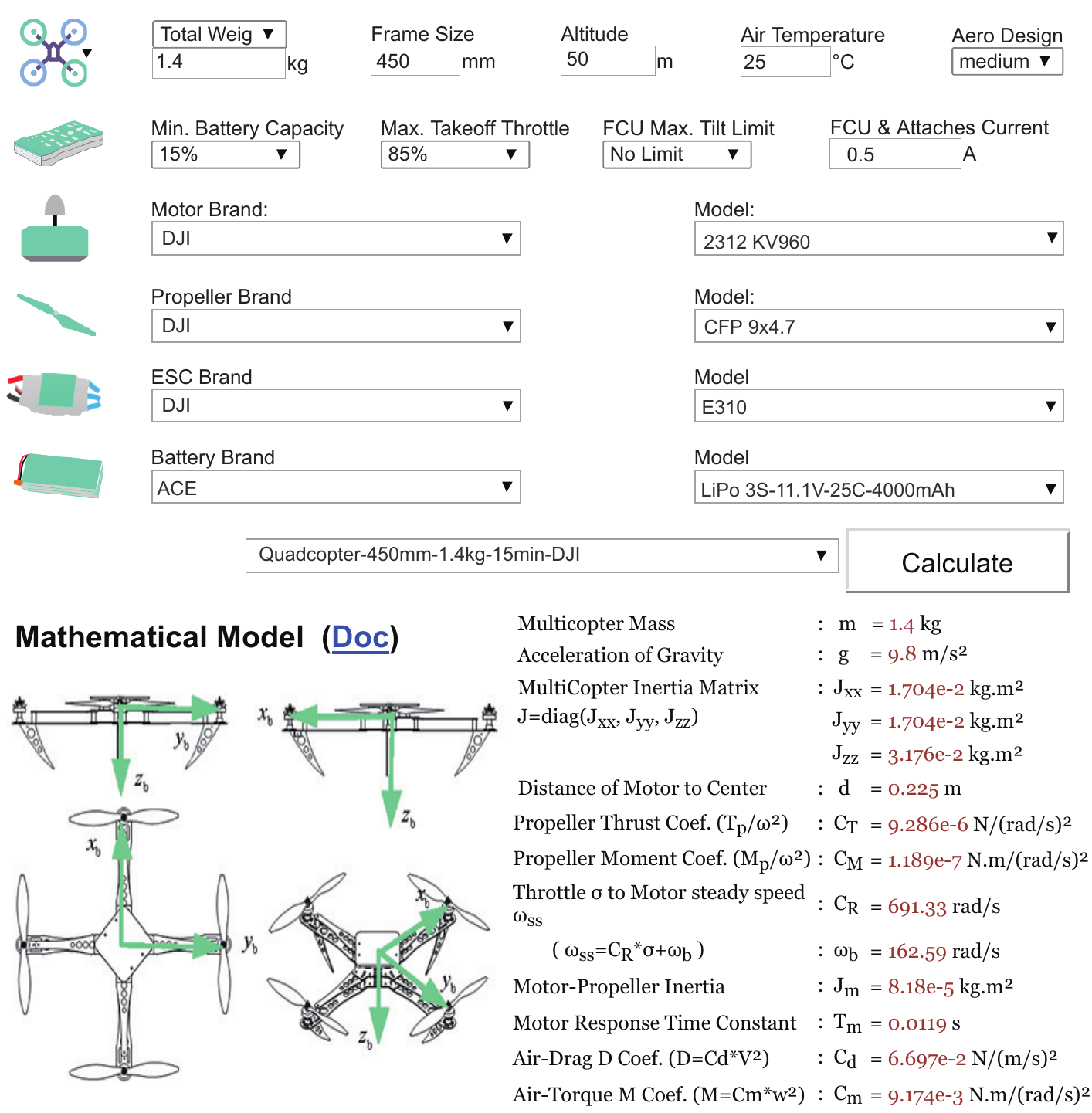}

\caption{Screenshot of the online toolbox flyeval.com.}
\label{Fig17}
\end{figure}

To take maximum advantage of the proposed test platform with MBD method,
a component model database is developed for the rapid development
of electric multicopters. The database covers the common products
on the market for multicopter propulsion systems with model parameters
obtained by their product specifications and experimental data. With
this model database, an online toolbox is released (URL: \uline{\url{https://flyeval.com/}})
based on our previous studies \cite{Shi2017,dai2018apractical,dai2018EFF}
for the automatic design and performance estimation of multicopter
UAVs, and a screenshot of the online toolbox is presented in Fig.\,\ref{Fig17}.
Users can select component products from the database to quickly assemble
a multicopter to estimate its flight performance and model parameters.
The toolbox has been released for more than two years, and the user
feedback indicates that it can significantly improve the multicopter
model development efficiency with decent simulation accuracy.

\begin{figure}
\centering \includegraphics[width=0.4\textwidth]{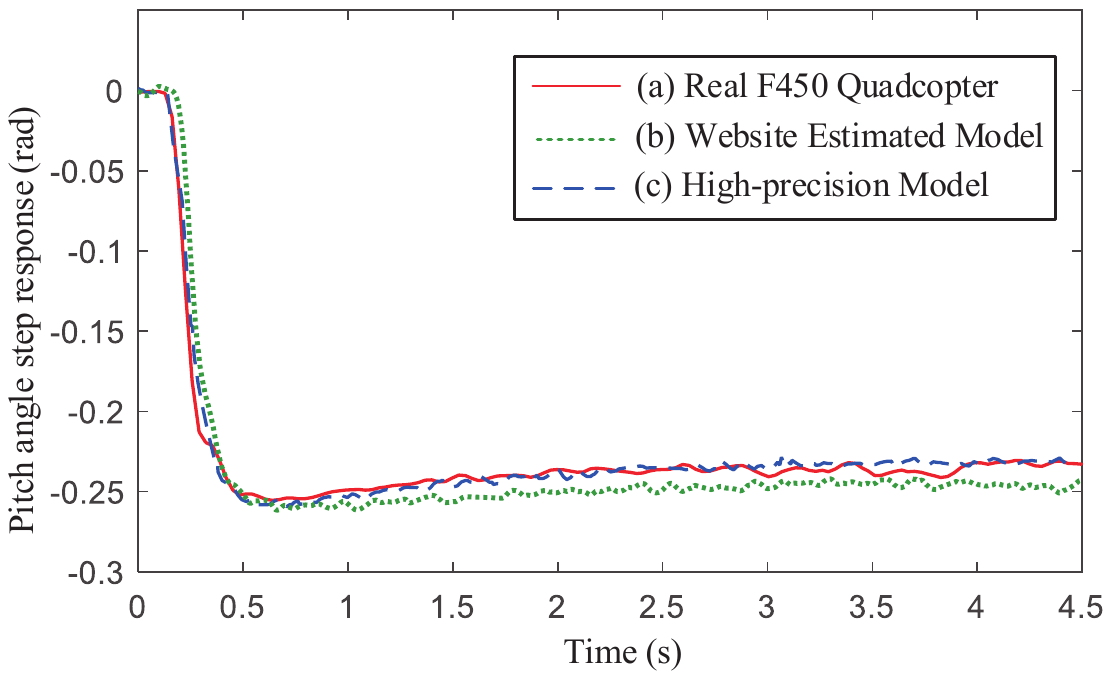}\caption{Level flight testing results for simulation validation.}
\label{Fig15-2}
\end{figure}

The level flight testing is an effective way to assess the simulation
accuracy of the whole simulation model because it is the joint effect
of all component models in the HIL test platform. Figs.\,\ref{Fig15-2}(a)(b)(c)
present the representative level flight testing results from a real
quadcopter, an estimated model from the toolbox in Fig.\,\ref{Fig17},
and a high-precision model calibrated with experimental data in Fig.\,\ref{Fig16-1}.
The quadcopter is commanded to step from hovering mode to level flight
mode (the desired pitch angle is 15 degree, i.e., 0.262 rad) at 0.2s.
The curve data are collected from the log data in Pixhawk after flight
tests. It can be observed from Fig.\,\ref{Fig15-2} that the high-precision
model curve almost coincides with the real experimental curve, and
the estimated model curve is slightly different from the experimental
curve, but the error is acceptable because it reveals most dynamic
and aerodynamic characteristics of the quadcopter.

\subsubsection{Algorithm Comparison}

\begin{figure}
\centering \includegraphics[width=0.45\textwidth]{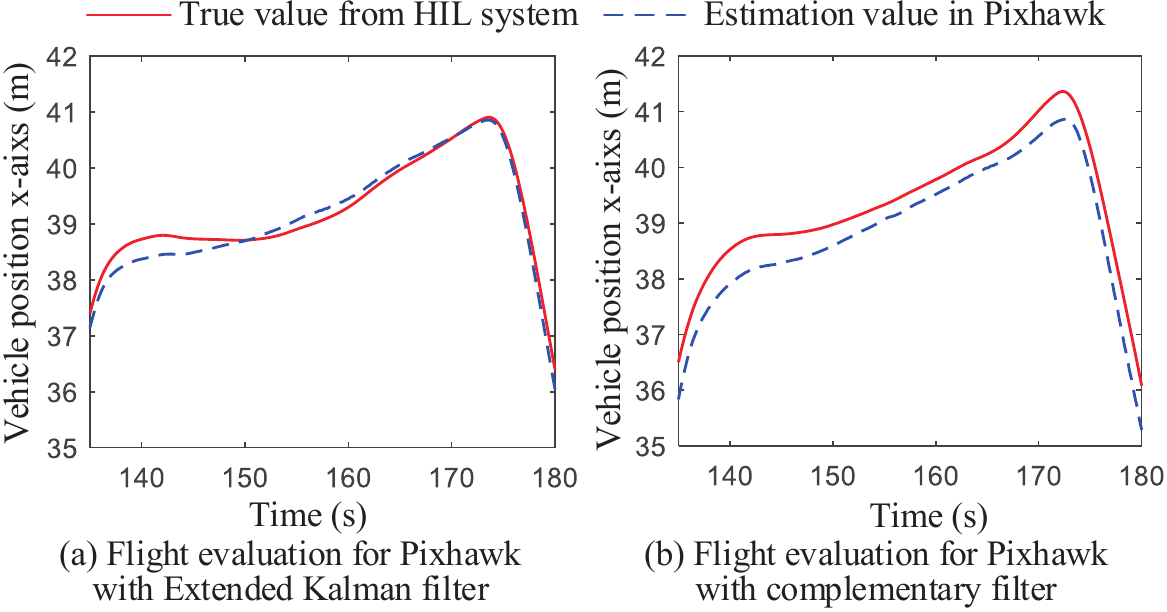}\caption{Comparing estimation performance of different filter algorithms in
turning flight stage.}
\label{Fig18-1}
\end{figure}

Another important advantage of the proposed HIL test platform is that
it can obtain the true states of the simulated vehicle, which is significant
for comparing performance difference of control algorithms. In Fig.\,\ref{Fig18-1},
two simulations are performed on the HIL test platform with two different
estimation filter algorithms in the Pixhawk autopilot. They are the
Extended Kalman filter algorithm in Fig.\,\ref{Fig18-1}(a) and the
complementary filter algorithm in Fig.\,\ref{Fig18-1}(b), respectively.
It can be observed from the result in Fig.\,\ref{Fig18-1} that the
extended Kalman filter has a better estimation effect than complementary,
which is consistent with the theoretical analysis. This conclusion
is hard to obtain through experiments because a higher more precise
external measuring devices (e.g., differential GPS or visual positioning
systems with centimeter-level precision) is required to measure the
true states of the vehicle. These external measuring devices are usually
expensive and restrained. For example, the differential GPS is easy
to be disturbed by flight environment factors, and its data frequency
is too low (usually 5Hz), and the visual positioning system cannot
be used outdoors. Therefore, the proposed test platform is a better
way to acquire the true states of the vehicle for comparative analysis
and performance assessment.

\subsubsection{Normal Testing}

\begin{figure}
\centering \includegraphics[width=0.45\textwidth]{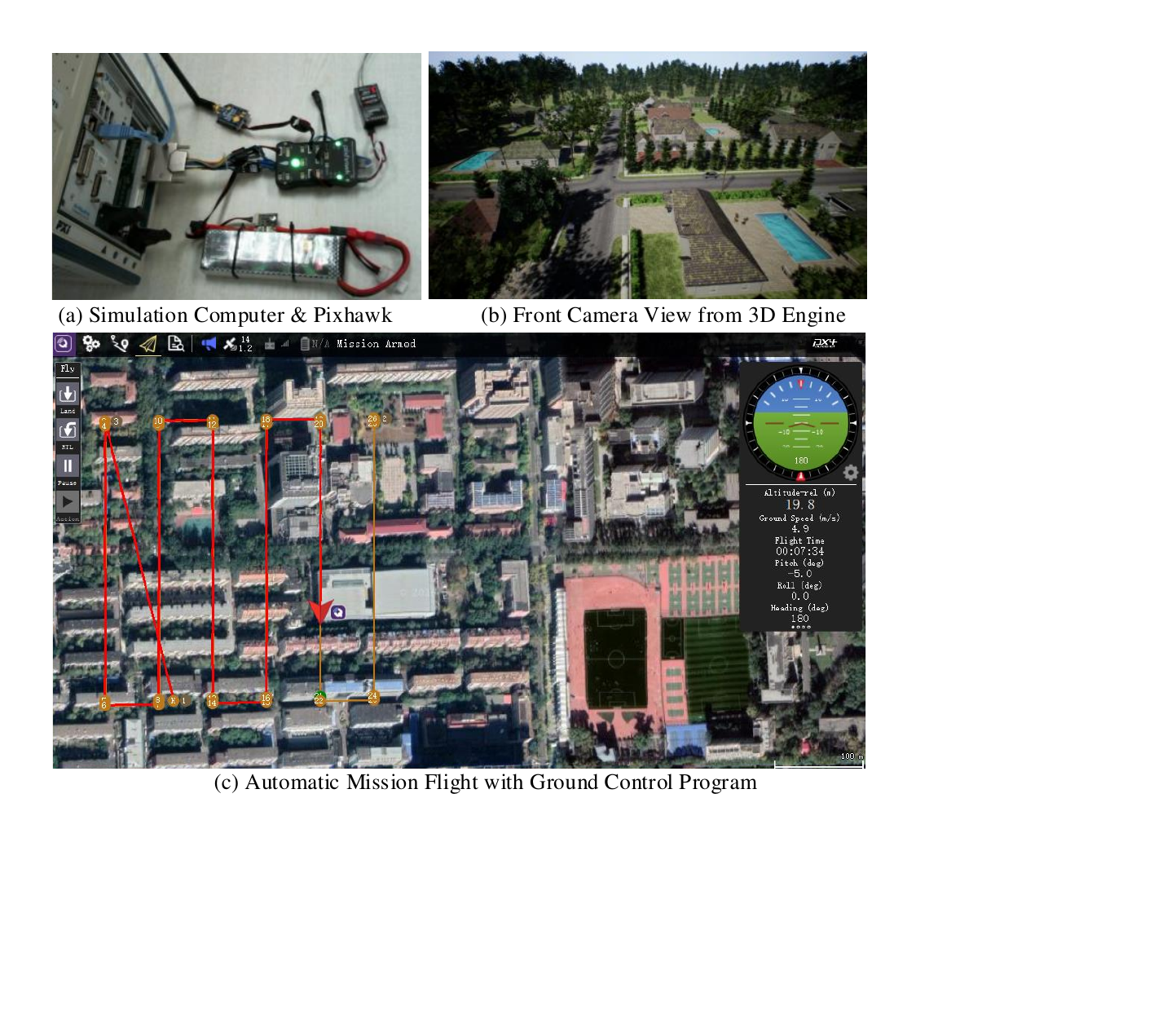}\caption{Autonomous mission flight testing with the proposed HIL test platform.}
\label{Fig18-2}
\end{figure}

The proposed HIL test platform makes it possible to comprehensively
test the control system only with computers, which is significant
in reducing cost and time relative to outdoor experiments. Fig.\,\ref{Fig18-2}
presents an autonomous mission flight test with the proposed HIL test
platform, which usually should be performed by outdoor flight tests
in traditional test methods. Besides, more comparative simulation
tests and experiment demonstrate that all flight tests (indoor or
outdoor, manual or automatic) can be tested on the HIL test platform
if the vehicle modeling is comprehensive and accurate enough.

\subsubsection{Automatic Safety Testing}

\begin{figure}
\centering \includegraphics[width=0.45\textwidth]{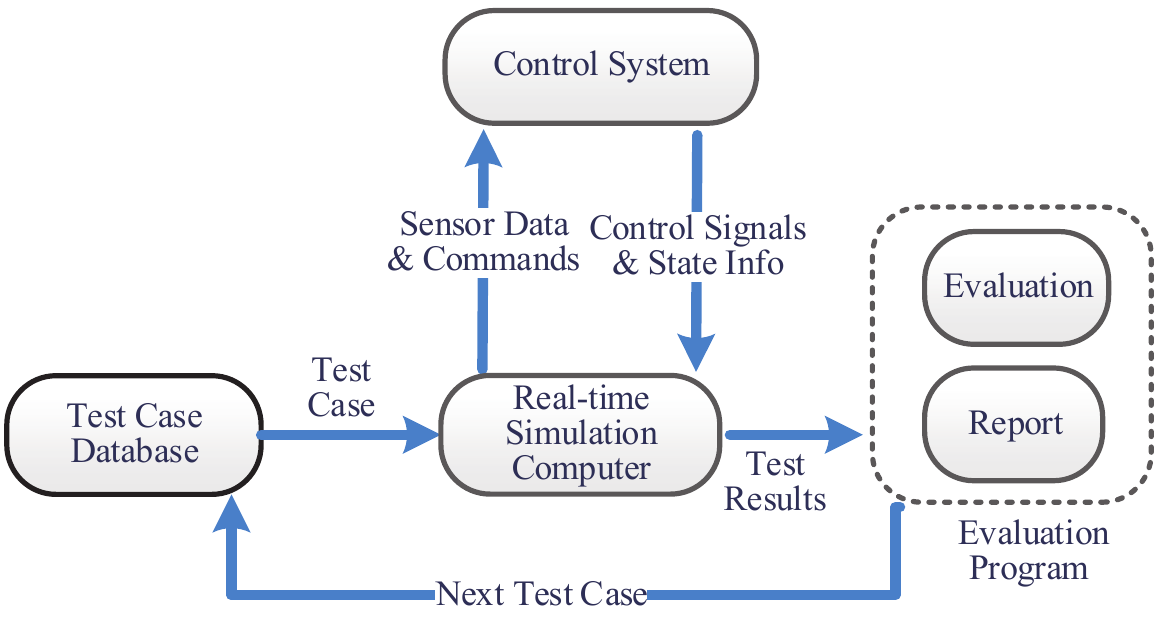}\caption{Automatic safety testing and validation framework.}
\label{Fig18}
\end{figure}

With the fault modes (e.g., sensor failure, wind disturbances, and
motor fault) being well modeled, the proposed HIL test platform can
also be applied to perform automatic safety testing for control systems.
We have developed an automatic safety testing framework (see Fig.\,\ref{Fig18})
based on the HIL test platform. First, a test case database should
be developed to store the vehicle command script, the fault injection
triggering time, the test stop time, and the desired vehicle flight
performance. In each testing case, the control system and the simulation
models are automatically reinitialized to default states, and then
the vehicle control system automatically controls the vehicle model
in the HIL test platform to the desired state. At this moment, a fault
case is injected to the HIL test platform to simulate a vehicle failure.
After a period of time, the vehicle and control system states are
automatically analyzed and recorded to a report. Finally, the next
testing case is tested in the same way until all the testing cases
are tested and evaluated. Some automatic test examples have been presented
in the attached video in the previous subsection. The test results
demonstrate that the platform can simulate the vehicle failure situations
realistically, and the test efficiency is significantly improved with
the automatic safety test method.

\section{Conclusion and Future Work}

\label{sec:5}

This paper presents a unified simulation and test platform, aiming
to significantly improve the development speed and safety level of
unmanned vehicle control systems. A unified modular modeling framework
is proposed by abstracting the common features of different types
of unmanned vehicles. The application examples demonstrate this framework
is efficient in developing a vehicle simulation model with compatibility
for the future safety assessment and certification standards. Another
key problem solved in the paper is to develop a HIL simulation test
platform to ensure simulation credibility. With the model-based design
method, the developing process of the simulation software can be automatized
and standardized, which ensures different developers can obtain the
same simulation software with credibility guaranteed by the automatic
code generation tools. With the FPGA-based real-time hardware-in-the-loop
simulation technology, the operating environment of the control algorithms
is guaranteed by using the same control system in both experiments
and simulations. After the credibility of the simulation test platform
is well guaranteed, we can focus on verifying and validating the simulation
models with quantitative assessment method proposed in our previous
work. The proposed test platform is applied to a multicopter control
system, where the accuracy and fidelity of the simulation testing
results are verified by comparing with experiments. The successful
applications present the advantages in the multicopter rapid prototyping,
estimation algorithm verification, autonomous flight testing, and
automatic safety testing with automatic fault injection and result
evaluation of unmanned vehicles.

Since the test platform can provide high-fidelity and credibility
simulation results, it will help to improve the training efficiency
of artificial intelligence algorithms. Besides, the airworthiness
for unmanned aerial vehicles require more formal, quantitative, and
efficient testing methods to assess the safety level, and we will
apply the proposed test platform for the safety assessment of unmanned
vehicle systems.

\bibliographystyle{IEEEtran}
\bibliography{IEEETRO}

\end{document}